\documentclass[10pt,a4paper]{article}

\usepackage[T1]{fontenc}
\usepackage[utf8]{inputenc}
\usepackage{lmodern}
\usepackage{microtype}

\usepackage[a4paper,
            top=1.5cm,
            bottom=1.5cm,
            left=1.5cm,
            right=1.5cm]{geometry}

\usepackage{amsmath,amssymb,bm}
\usepackage{graphicx}
\usepackage{booktabs}
\usepackage{tabularx}
\usepackage{subcaption}
\usepackage{placeins}
\usepackage{xcolor}
\usepackage{enumitem}
\usepackage{authblk}

\usepackage[authoryear,round,sort]{natbib}
\newcommand{\SU}{\mathrm{SU}}

\newcommand{\BOne}{B_{1}}
\newcommand{\Fgate}{F_{\mathrm{gate}}}

\newcommand{\rev}[1]{{\color{black}#1}}
\usepackage[colorlinks=true,
            citecolor=blue,
            linkcolor=blue,
            urlcolor=blue]{hyperref}

\providecommand{\doi}[1]{}
\renewcommand{\doi}[1]{%
  \href{https://doi.org/#1}{https://doi.org/#1}%
}
\title{\textbf{Risk-aware neural pulse compilation of \rev{continuously parameterized single-qubit} gates
with optimal-control benchmarking and NMR hardware validation}}

\author[3,4]{Arash Fath Lipaei}
\author[1]{Ebrahim Khaleghian\thanks{Corresponding author: \href{mailto:ebrahim.khaleghian@gmail.com}{ebrahim.khaleghian@gmail.com}}}
\author[5]{Gani Göral}
\author[6]{Zidong Lin}
\author[3,4]{Selin Aslan}
\author[1,2]{Özgür E. Müstecaplıoğlu}

\affil[1]{Department of Physics, Koç University, Sarıyer, 34450 Istanbul, Türkiye}

\affil[2]{TÜBİTAK Research Institute for Fundamental Sciences (TBAE), 41470, Gebze, Türkiye}

\affil[3]{Computational Sciences and Engineering Department, Koç University, Sarıyer, 34450, Istanbul, Türkiye}

\affil[4]{Department of Mathematics,
Koç University, Sarıyer, 34450, Istanbul, Türkiye}

\affil[5]{Mechanical Engineering Department,
Middle East Technical University, 06800, Ankara, Türkiye}

\affil[6]{Shenzhen SpinQ Technology Co., Ltd., 518043, Shenzhen, China}

\date{}

\begin{document}

\maketitle

\begin{center}
\textit{Arash Fath Lipaei and Ebrahim Khaleghian contributed equally to this work and share first authorship.}
\end{center}

\begin{abstract}
Pulse-level compilation can reduce the depth and calibration overhead associated with decomposing continuously parameterized quantum operations into
fixed gate libraries, but conventional optimal control requires a new optimization for each target and can be sensitive to hardware uncertainty. We introduce a
physics-informed neural compilation framework that learns a reusable map from \rev{continuously parameterized single-qubit gate targets} directly to experimentally admissible control
waveforms. The compiler is trained end-to-end through differentiable quantum dynamics without precomputed optimal-pulse labels, and its robustness is
improved by first diagnosing physically relevant failure modes and then applying tail-risk-aware fine-tuning. On a three-qubit liquid-state nuclear magnetic resonance (NMR) model, the learned compiler generalizes across Haar-distributed single-qubit targets with mean nominal fidelity above 0.99 and generates new pulses in submillisecond inference time. Matched benchmarks reveal complementary computational advantages: robust gradient ascent pulse engineering
(GRAPE) achieves the highest per-target fidelity, whereas the trained neural compiler amortizes the optimization cost across the target family, enabling substantially faster online pulse generation and effective initialization of subsequent target-specific refinement. Risk-aware training substantially broadens tolerance
to radio-frequency (RF)-amplitude errors, including perturbations beyond the
training range. Experiments on a three-spin NMR processor \rev{reproduce the predicted broadening of the RF-tolerance response, although the absolute experimental correlations are lower than the simulated values}. These results establish a general
workflow combining amortized machine learning, physical sensitivity diagnosis,
risk-aware control, optimal-control benchmarking, and hardware testing, providing a route toward fast and robust pulse-level compilation for repeatedly varying
quantum operations.

\end{abstract}

\noindent\textbf{Keywords:}
machine learning;
quantum optimal control;
neural pulse compilation;
robust quantum control;
differentiable quantum control;
NMR quantum computing

\section{Introduction}\label{sec:introduction}

Quantum algorithms are ultimately implemented as controlled physical dynamics, not as abstract unitary matrices. The translation from a logical target to experimentally admissible fields therefore sits at the interface between quantum compilation, calibration, and optimal control \citep{Koch2022QuantumOptimalControl,Clinton2021HamiltonianSimulation}. This interface is especially explicit in liquid-state nuclear magnetic resonance (NMR), where coupled nuclear spins evolve under accurately characterized drift interactions and shaped radio-frequency (RF) fields, making pulse design inseparable from the Hamiltonian model \citep{Vandersypen2005NMRTechniques,Gershenfeld1997BulkSpinResonance,Cory1997EnsembleQuantumComputing,Khaneja2005OptimalControl}. NMR is consequently both a mature setting for quantum control and a useful experimental testbed for studying learning-based pulse synthesis under controlled model mismatch. More broadly, machine-learning methods for quantum estimation and control now span neural-network-based control, differentiable optimization, evolutionary methods, robust-control strategies, and reinforcement learning, reflecting increasing interest in using learned models to complement conventional optimal-control techniques \citep{Ma2025MLQuantumControl}.

A conventional quantum compiler maps target operations to a discrete universal gate library, after which calibrated hardware primitives implement the resulting circuit \citep{Barenco1995,Dawson2005SolovayKitaev}. This separation is effective when a small reusable gate alphabet is sufficient, but it is less natural for continuously parameterized transformations. Variational quantum algorithms, adaptive circuits, and pulse-driven ans\"atze repeatedly request rotations whose parameters change during execution or optimization \citep{Cerezo2021,Kandala2017,Grimsley2019,Ibrahim2022}. Decomposing each such target into a fixed basis can add depth, while re-running a high-dimensional pulse optimizer for every target incurs a different cost: compilation latency. Experimental work on superconducting hardware has likewise shown that arbitrary operations can be compiled directly at the pulse level, avoiding potentially long decompositions into primitive gates and demonstrating the practical relevance of direct optimal-control synthesis beyond NMR \citep{Cho2024DirectPulseCompilation}. At the complementary circuit level, learning-based synthesis has also been pursued using reinforcement learning to search for low-depth, low-gate-count circuits, illustrating the broader role of machine learning across multiple layers of the quantum compilation stack \citep{Giordano2026CircuitSynthesis}. Recent software efforts have further begun to expose pulse-level modeling directly within quantum-machine-learning workflows, motivated by the additional expressivity available below the gate abstraction \citep{Franz2026PulseSoftware}. The resulting problem is to learn a reusable mapping from target-unitary parameters to high-fidelity controls across the continuous target family.

The closest prior literature already establishes several ingredients. Neural optimization of continuous gate families was demonstrated by \citet{Sauvage2022}, and \citet{Preti2022} learned parameter-dependent pulse classes through both supervised and direct differentiable optimization, including uncertain physical parameters. Differentiable programming has also been used to optimize neural-network control policies by propagating gradients jointly through the network and the governing quantum dynamics, establishing a closely related model-based route to learning controls without treating an external optimizer as a black box \citep{Schafer2020Differentiable}. More recently, deep neural networks have been used to learn pulse-parameter-to-fidelity relations for selected quantum gates and thereby accelerate pulse-waveform optimization \citep{Fillingham2025PulseWaveforms}; unlike such surrogate optimization approaches, the compiler studied here maps the continuously parameterized target itself directly to the complete time-dependent control waveform. \rev{A previous NMR study learned a gate-to-pulse map from gradient ascent pulse engineering (GRAPE)-generated data and implemented neural-network-generated single-qubit controls experimentally} \citep{PrevWork}. \rev{Relative to that supervised gate-to-pulse study, the present work adds (i) label-free end-to-end training through the simulated dynamics, (ii) per-target CVaR-based risk-aware fine-tuning after a sensitivity diagnosis, (iii) matched nominal, robust, and Neural-Network-seeded GRAPE benchmarks, and (iv) an NMR hardware test focused on the diagnosed RF-gain error channel.} Robust quantum control through sampled uncertainty ensembles also predates neural pulse compilation: sampling-based learning control was developed to train controls over uncertain model realizations and test them on additional uncertainty samples \citep{Dong2015SamplingLearning}, and the same general framework was subsequently applied to robust universal quantum-gate pulse generation \citep{Dong2016RobustPulses}. \rev{RF-amplitude-insensitive pulse design is also established in NMR: broadband and passband composite pulses compensate RF inhomogeneity, while optimal-control-designed broadband excitation and inversion pulses address similar errors for individual operations} \citep{Wimperis1994,Kobzar2004}. \rev{The distinction here is not the invention of RF-robust pulses, but learning to produce them rapidly across a continuously parameterized target family.} Modern robust quantum control now encompasses a broad range of ensemble, uncertainty-aware, worst-case, and learning-assisted techniques \citep{Weidner2025RobustQuantumControl}. Robust neural pulse design also predates the present study, smooth noise-insensitive pulses have been obtained with physics-informed neural networks \citep{Gungordu2022, Weidner2025RobustQuantumControl}, supervised learning has been applied to robust composite pulses \citep{Shi2024}, and \citet{Kuzmanovic2026} recently demonstrated a detuning-robust neural-ODE $\pi/2$ pulse on superconducting hardware. Recent learning-based alternatives have further explored meta-reinforcement learning for universal-gate control under uncertain environments \citep{Zhang2025MetaQctrl}, diffusion-based reinforcement learning for pulse-level robustness against control and Hamiltonian uncertainties \citep{Zhang2025DiffusionRL}, and reinforcement learning from model-generated demonstrations to mitigate model bias while retaining improved sample efficiency \citep{Li2025RLfD}. Conditional Value-at-Risk (CVaR) itself is likewise established, including in variational quantum optimization \citep{Barkoutsos2020}. Relative to these closest precedents, our question is: can a reusable continuous-family compiler be stress-diagnosed, tail-risk-robustified, compared fairly with strong per-target robust optimal control, and then tested experimentally under the same identified failure mode?

We first freeze the nominal compiler and diagnose its response to structured, experimentally motivated perturbations. The resulting sensitivity profile determines which physical error channel is emphasized during robust fine-tuning. Related machine-learning work has emphasized that experimentally observed noise can itself contain device-specific structure that can be learned and classified, motivating diagnosis of the relevant physical error landscape rather than treating hardware imperfections as undifferentiated noise \citep{Martina2022NoiseFingerprint}. Among the channels examined, quasi-static global RF-gain error---an effective $\BOne$ calibration/inhomogeneity error---produces the dominant degradation. We then train a second compiler with a risk-sensitive objective based on CVaR. CVaR, originally developed as a coherent tail-risk measure \citep{Acerbi2002,Rockafellar2000,RockafellarUryasev2002}, assigns explicit weight to the adverse tail of a loss distribution. In our setting the loss is gate infidelity under sampled physical perturbations, so the objective penalizes pulse families whose performance is acceptable on average but fails sharply for unfavorable yet plausible realizations. The resulting nominal-to-robust transition is therefore a controlled trade-off: some zero-error fidelity is sacrificed in exchange for a substantially wider high-fidelity region.

A central methodological point is that a neural compiler and a target-specific optimizer solve related but different computational problems. Gradient-ascent pulse engineering (GRAPE) is a powerful offline method for finding high-quality controls for a specified target \citep{Khaneja2005OptimalControl,Koch2022QuantumOptimalControl}. A trained neural compiler instead pays a large optimization cost once and then evaluates a learned map for new targets. In this sense, the learned mapping may be viewed as an amortized form of quantum optimal control: optimization effort is transferred from repeated target-specific searches into the training of a shared target-conditioned model. We evaluate nominal and risk-aware neural compilers against standalone GRAPE, robust GRAPE, and neural-network-seeded GRAPE under matched physical constraints, reporting both control quality and compilation time. The purpose is to characterize the quality--latency--robustness trade space created by amortization.

We next examine whether the numerically identified RF-gain response survives hardware transfer. Nominal and risk-aware pulses are applied on a three-qubit $^{19}\mathrm{F}$ liquid-state NMR processor, followed by an RF-gain sweep around a Hadamard target. State tomography is reconstructed with one receiver-phase calibration obtained from a pseudo-pure-state reference and then \rev{kept fixed} for all perturbed experiments. Simulation and experiment are compared using the same normalized Hilbert--Schmidt correlation in deviation-matrix space. This protocol tests the specific physical prediction produced by robust training. A nominal pulse should peak near calibration and degrade rapidly under gain error, while a risk-aware pulse should exhibit a broader fidelity plateau.

Pulse compilation across a continuous family of quantum operations creates a natural setting for amortized machine learning. Physics-based optimization is performed during training across the target family, producing a reusable model that generates controls for new target operations at neural-network inference speed. This capability is particularly useful in variational algorithms, adaptive protocols, and calibration procedures that repeatedly request different parameterized operations. Experimentally diagnosed uncertainties can also be incorporated during training, allowing the learned control map to combine rapid pulse generation with tolerance to relevant hardware imperfections. Our results show how these elements can be integrated into a single control framework that provides fast, reusable, and physically robust pulse compilation across a continuous family of \rev{single-qubit operations}.

The study combines five linked stages: label-free end-to-end training
over a Haar-distributed SU(2) family; sensitivity diagnosis using the nominal compiler;
per-gate CVaR fine-tuning with in- and out-of-training-range tests; matched nominal GRAPE,
robust GRAPE, and NN-seeded GRAPE benchmarks that separate amortized inference from
per-target optimization; and NMR tomography of the resulting nominal–robust RF-gain response.
This sequence allows the numerical robustness mechanism, its computational trade-offs, and its
hardware manifestation to be evaluated within one control model.

\section{Continuous neural quantum pulse compilation}\label{sec:compiler}

\subsection{Three-spin control model and differentiable propagation}

We model the processor as three spin-$1/2$ nuclei driven by one global transverse RF channel.  Let $X_i$, $Y_i$, and $Z_i$ denote Pauli operators acting on spin $i$ and identity on the other spins ~\citep{Vandersypen2005NMRTechniques,Khaneja2005OptimalControl,PrevWork}.  In the rotating-frame convention used by the numerical compiler, the drift Hamiltonian is
\begin{equation}
H_0=\pi\sum_{i=1}^{3}\nu_i Z_i
+\frac{\pi}{2}\sum_{1\leq i<j\leq3}J_{ij}
\left(X_iX_j+Y_iY_j+Z_iZ_j\right),
\label{eq:drift}
\end{equation}
with $(\nu_1,\nu_2,\nu_3)=(-921,40.75,700)\,\mathrm{Hz}$ and $(J_{12},J_{13},J_{23})=(-128,48.8,68.2)\,\mathrm{Hz}$. Equation~\eqref{eq:drift} is the pair-interaction model used consistently in the nominal training, robustness analysis, optimal-control benchmarks, and hardware-transfer simulations. Collective transverse operators are
\begin{equation}
H_x=X_1+X_2+X_3,\qquad H_y=Y_1+Y_2+Y_3,
\end{equation}
and the $k$th piecewise-constant control slice is
\begin{equation}
H_k=H_0+\pi\left(u_{x,k}H_x+u_{y,k}H_y\right).
\label{eq:Hk}
\end{equation}
The coefficients $u_{x,k}$ and $u_{y,k}$ are expressed in Hz-equivalent units.  The pulse has $N=300$ slices of duration $\Delta t=35\,\mu\mathrm{s}$, giving a total duration $T=N\Delta t=10.5\,\mathrm{ms}$.  These values are the timing convention used by the checkpoints and by the later robustness, benchmark, and hardware-transfer phases.

The slice propagators and time-ordered total propagator are
\begin{equation}
U_k=\exp[-i\Delta t H_k],\qquad
U(\mathbf u)=U_NU_{N-1}\cdots U_1,
\label{eq:propagator}
\end{equation}
where $\mathbf u\in\mathbb R^{N\times2}$ collects the two RF quadratures.  Matrix exponentials and the ordered product are evaluated inside the computational graph, so automatic differentiation propagates the gate-fidelity gradient through the same dynamics used for evaluation.  No pulse labels enter the nominal training loss.

\subsection{Haar-distributed continuous target family}

The compiler acts on the first spin while leaving the remaining two unchanged.  The final model uses a unit quaternion
\begin{equation}
\mathbf q=(q_0,q_x,q_y,q_z)\in S^3,
\qquad \|\mathbf q\|_2=1,
\end{equation}
to represent an arbitrary element of $\SU(2)$,
\begin{equation}
U_2(\mathbf q)=q_0 I_2-i(q_xX+q_yY+q_zZ).
\label{eq:quaternionSU2}
\end{equation}
The full-register target is
\begin{equation}
U_{\mathrm{tar}}(\mathbf q)=U_2(\mathbf q)\otimes I_2\otimes I_2.
\label{eq:lift}
\end{equation}
During training, four independent standard-normal variates are drawn and normalized to unit length; the resulting distribution is uniform on $S^3$ and therefore induces Haar-uniform sampling of $\SU(2)$ up to the physically irrelevant double-cover sign.  The quaternion representation avoids coordinate singularities in the target distribution while retaining only four real input features \citep{Hamermesh1962}.

For $d=8$, performance is measured with the global-phase-insensitive trace fidelity
\begin{equation}
\Fgate(U_{\mathrm{tar}},U)
=\frac{1}{d^2}\left|\operatorname{Tr}\!\left(U_{\mathrm{tar}}^{\dagger}U\right)\right|^2,
\qquad d=8.
\label{eq:unitary_fidelity}
\end{equation}

\subsection{Neural map, bounded controls, and nominal training}

The compiler is a multilayer perceptron $f_\theta:S^3\rightarrow\mathbb R^{N\times2}$ with ten hidden layers of width $256$, GELU nonlinearities, dropout probability $0.25$, and a $600$-component output head.  The final architecture contains $747{,}608$ trainable parameters.  If $\mathbf z_k\in\mathbb R^2$ is the unbounded output for slice $k$, the physical control is produced by the smooth radial map
\begin{equation}
\mathbf u_k
=u_{\max}\frac{\mathbf z_k}{\sqrt{1+\|\mathbf z_k\|_2^2}},
\qquad u_{\max}=50\,\mathrm{kHz},
\label{eq:radial_bound}
\end{equation}
which guarantees $\|\mathbf u_k\|_2<u_{\max}$ without component-wise clipping.  \rev{The $50\,\mathrm{kHz}$ quantity is a numerical optimization bound, not a separately established instrument-amplitude limit.}

For a batch $\{\mathbf q_b\}_{b=1}^{B}$, the nominal objective is
\begin{align}
\mathcal L_{\mathrm{nom}}
&=1-\frac1B\sum_{b=1}^{B}\Fgate^{(b)}
+\lambda_{\mathrm{s}}\mathcal R_{\mathrm{s}}
+\lambda_{\mathrm{p}}\mathcal R_{\mathrm{p}},\label{eq:nominal_loss}\\
\mathcal R_{\mathrm{s}}
&=\frac{1}{B(N-1)}\rev{\sum_{b=1}^{B}\sum_{k=1}^{N-1}}
\frac{\|\mathbf u_{b,k+1}-\mathbf u_{b,k}\|_2^2}{u_{\max}^2},\\
\mathcal R_{\mathrm{p}}
&=\frac{1}{BN}\rev{\sum_{b=1}^{B}\sum_{k=1}^{N}}\frac{\|\mathbf u_{b,k}\|_2^2}{u_{\max}^2},
\end{align}
with $\lambda_{\mathrm{s}}=10^{-4}$ and $\lambda_{\mathrm{p}}=10^{-6}$.  AdamW is used with initial learning rate $5\times10^{-4}$, weight decay $10^{-3}$, cosine decay to $5\%$ of the initial learning rate, gradient-norm clipping at $1$, and batch size $32$.  Training runs for $50{,}000$ steps; model selection is based on mean fidelity on a fixed set of $512$ unseen Haar targets.  The nominal checkpoint used by all subsequent phases was selected at step $48{,}600$. Table~\ref{tab:nominal_hparams} collects the architecture and optimization settings.

\begin{table}[t]
\centering
\caption{\rev{Nominal} neural-compiler architecture and training hyperparameters.  ``LR floor'' denotes the final cosine-decay fraction of the initial learning rate.}
\label{tab:nominal_hparams}
\footnotesize
\renewcommand{\arraystretch}{1.08}
\setlength{\tabcolsep}{4.2pt}
\begin{tabular}{ll@{\hspace{1.0em}}ll}
\toprule
Architecture/control & Value & Optimization & Value\\
\midrule
Input representation & $\mathbf q\in S^3$  & Optimizer & AdamW\\
Hidden stack & $10\times256$ GELU & Initial LR & $5\times10^{-4}$\\
Dropout & $0.25$ & Weight decay & $10^{-3}$\\
Output & $300\times2=600$ & Batch size & $32$\\
Trainable parameters & $747{,}608$ & Training steps & $50{,}000$\\
Slices / duration & $300$ / $10.5\,\mathrm{ms}$ & LR floor & $0.05\,\mathrm{LR}_0$\\
Slice width & $35\,\mu\mathrm{s}$ & Gradient clip & $1.0$\\
Numerical amplitude cap & $50\,\mathrm{kHz}$ & $(\lambda_s,\lambda_p)$ & $(10^{-4},10^{-6})$\\
Validation set & $512$ Haar targets & Validation cadence & $100$ steps\\
\bottomrule
\end{tabular}
\end{table}

\section{Structured uncertainty and risk-aware robustification}\label{sec:robust}

\subsection{Nominal-model stress diagnosis}

Robustification is motivated by a separate \rev{nominal-model} stress test.  Let a scenario $\xi$ contain a common carrier-field offset $\gamma\delta B_0$, independent chemical-shift perturbations $\delta\nu^{cs}_i$ , coupling perturbations $\delta J_{ij}$, RF gain g, phase-reference error $\phi$, clock-scale error c (ppm), and slice timing jitter $\delta t_k$, where $\gamma$ is the
gyromagnetic ratio. The scenario-dependent drift parameters are
\begin{equation}
\widetilde\nu_i=\nu_i+\frac{\gamma}{2\pi}\delta B_0+\delta\nu^{cs}_i,
\qquad
\widetilde J_{ij}=J_{ij}+\delta J_{ij},
\label{eq:drift_pert}
\end{equation}
and gain/phase errors transform the two commanded quadratures as
\begin{equation}
\begin{bmatrix}\widetilde u_{x,k}\\[1mm]\widetilde u_{y,k}\end{bmatrix}
=g
\begin{bmatrix}\cos\phi&-\sin\phi\\ \sin\phi&\cos\phi\end{bmatrix}
\begin{bmatrix}u_{x,k}\\[1mm]u_{y,k}\end{bmatrix}.
\label{eq:control_pert}
\end{equation}
Timing is perturbed according to
\begin{equation}
\widetilde{\Delta t}_k
=\Delta t\left(1+c\times10^{-6}\right)+\delta t_k.
\label{eq:time_pert}
\end{equation}
Importantly, $g$, $\phi$, $\frac{\gamma}{2\pi}\delta B_0$, $\delta\nu^{cs}_i$, $\delta J_{ij}$, and $c$ are \emph{quasi-static over the full 300-slice pulse} for one scenario; they are not redrawn independently at every time slice.  Only the small timing-jitter term varies slice-by-slice.  This distinction is essential for interpreting the RF-gain variable as a $B_1$ calibration/inhomogeneity error.

\begin{figure}[t]
\centering
\begin{subfigure}[t]{0.485\linewidth}\centering
\includegraphics[width=\linewidth]{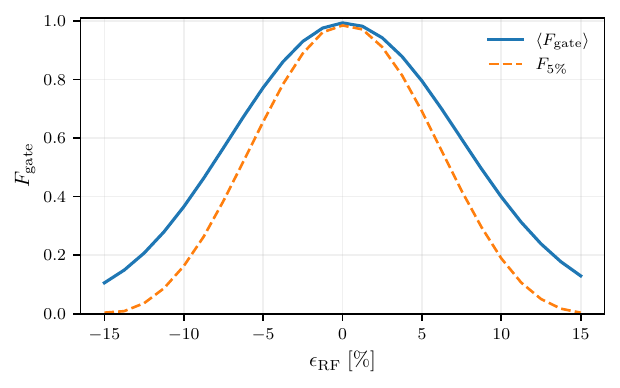}
\caption{Global RF/$B_1$ gain.}
\end{subfigure}\hfill
\begin{subfigure}[t]{0.485\linewidth}\centering
\includegraphics[width=\linewidth]{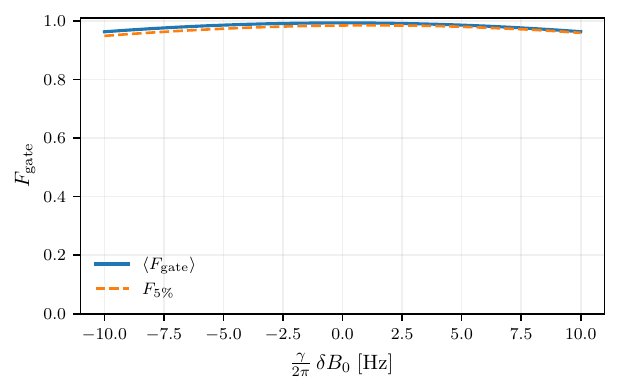}
\caption{Common $\frac{\gamma}{2\pi}\delta B_0$ offset.}
\end{subfigure}

\begin{subfigure}[t]{0.485\linewidth}\centering
\includegraphics[width=\linewidth]{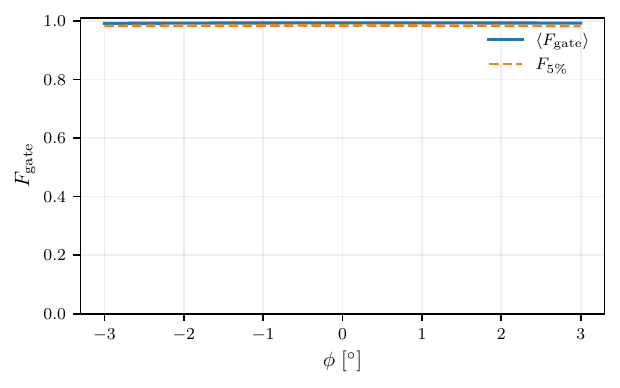}
\caption{Static RF phase bias.}
\end{subfigure}\hfill
\begin{subfigure}[t]{0.485\linewidth}\centering
\includegraphics[width=\linewidth]{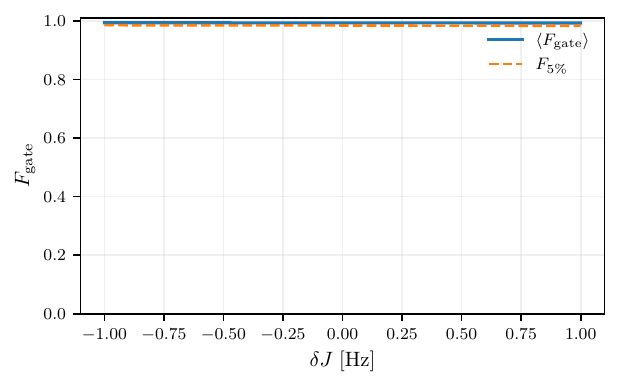}
\caption{Coupling-magnitude shift.}
\end{subfigure}
\caption{Sensitivity diagnosis before robust training.  Each panel sweeps one physical channel while holding the others nominal; solid and dashed curves are the target-averaged fidelity and fifth percentile, respectively.  The much sharper RF-gain response motivates the subsequent risk-aware redesign.}
\label{fig:stress_channels}
\end{figure}

The routine joint envelope uses $|\frac{\gamma}{2\pi}\delta B_0|\le2\,\mathrm{Hz}$, $|\delta\nu^{cs}_i|\le2\,\mathrm{Hz}$, $|\delta J_{ij}|\le0.25\,\mathrm{Hz}$, $|g-1|\le0.05$, $|\phi|\le1^\circ$, $|c|\le2\,\mathrm{ppm}$, and Gaussian timing jitter with $10\,\mathrm{ps}$ RMS.  A wider conservative stress envelope uses $6\,\mathrm{Hz}$, $6\,\mathrm{Hz}$, $1\,\mathrm{Hz}$, $15\%$, $3^\circ$, $5\,\mathrm{ppm}$, and $50\,\mathrm{ps}$ RMS, respectively.  One-dimensional sweeps and joint Sobol scenarios were applied to the nominal checkpoint, with no retraining.  Figure~\ref{fig:stress_channels} shows four representative channels using both the target-averaged mean fidelity and the fifth percentile.  The contrast is pronounced: at $\epsilon_{\rm RF}=\pm15\%$ the mean drops to $0.105/0.129$ and the fifth percentile is approximately $0.0035$, whereas a $\pm10\,\mathrm{Hz}$ common-$B_0$ shift still leaves the mean near $0.963$.  Static phase bias through $\pm3^\circ$ and a common coupling-magnitude shift through $\pm1\,\mathrm{Hz}$ have still smaller effects. The remaining tested channels produced broad fidelity plateaus over the examined ranges and were therefore not selected as the primary robust-training axis. This diagnosis identifies RF gain as the uncertainty axis emphasized in robust training and in the subsequent hardware experiment.

\subsection{Empirical CVaR objective}

For target $b$ and uncertainty scenario $s$, define the infidelity
\begin{equation}
\ell_{b,s}(\theta)=1-\Fgate\!\left(U_{\mathrm{tar},b},U(f_\theta(\mathbf q_b);\xi_s)\right).
\end{equation}
We use upper-tail CVaR with a \emph{tail-mass} convention $\alpha\in(0,1]$.  In population form this is the Rockafellar--Uryasev functional \citep{Rockafellar2000,RockafellarUryasev2002,Acerbi2002,Shapiro2014}
\begin{equation}
\rho_{\alpha}(L)
=\min_{\eta\in\mathbb R}\left[\eta+\frac{1}{\alpha}\,\rev{\mathbb E[(L-\eta)_+]}\right].
\label{eq:ru_cvar}
\end{equation}

\rev{Here, $L$ denotes the random gate infidelity
under the sampled hardware uncertainties,
$\eta\in\mathbb{R}$ is an auxiliary loss threshold,
$(x)_+=\max(x,0)$, and $\mathbb{E}$ denotes
expectation over the uncertainty distribution.
The quantity $\rho_\alpha(L)$ measures the
average loss in the worst $\alpha$ fraction of
uncertainty realizations.}

\rev{The implementation uses a finite-sample top-$k$ tail-mean estimator.}  If $S$ scenarios are evaluated for a gate and $k=\lceil\alpha S\rceil$, let $\mathcal W_k(b)$ index the $k$ largest values of $\ell_{b,s}$.  Then
\begin{equation}
\widehat\rho_{\alpha,b}
=\frac1k\sum_{s\in\mathcal W_k(b)}\ell_{b,s}.
\label{eq:emp_cvar}
\end{equation}
\rev{For the training configuration $\alpha=0.20$ and $S=10$, the estimator agrees exactly with empirical CVaR because $\alpha S=2$. For noninteger $\alpha S$, the unweighted top-$k$ mean is an approximation to empirical CVaR.}
CVaR is therefore computed \emph{separately for each target gate} before averaging across the batch; pooling losses from different targets would conflate target difficulty with scenario risk.

The robust network is warm-started from the nominal checkpoint and optimized with
\begin{equation}
\mathcal L_{\mathrm{rob}}
=\lambda_0\underbrace{\mathbb E_b[1-F_{b,0}]}_{\mathcal L_0}
+\lambda_\mu\underbrace{\mathbb E_{b,s}[\ell_{b,s}]}_{\mathcal L_\mu}
+\lambda_{\rm tail}\underbrace{\mathbb E_b[\widehat\rho_{\alpha,b}]}_{\mathcal L_{\rm tail}}
+\lambda_{\mathrm{s}}\mathcal R_{\mathrm{s}}+\lambda_{\mathrm{p}}\mathcal R_{\mathrm{p}},
\label{eq:robust_loss}
\end{equation}
\rev{Here, $F_{b,0}$ denotes the fidelity of target $b$ under nominal (uncertainty-free) conditions, while $\mathbb E_b$ and $\mathbb E_{b,s}$ denote averages over target gates and over both target gates and sampled uncertainty scenarios, respectively. The loss weights are $(\lambda_0,\lambda_\mu,\lambda_{\rm tail})=(1,1,1.5)$, with $\alpha=0.20$.}  Each training microbatch contains eight random joint scenarios plus the exact negative and positive RF-gain boundaries; hence $S=10$ and the empirical tail term averages the two worst scenarios for each gate.  Random non-RF uncertainties remain inside the routine envelope above.

\begin{figure}[t]
\centering
\includegraphics[width=0.5\linewidth]{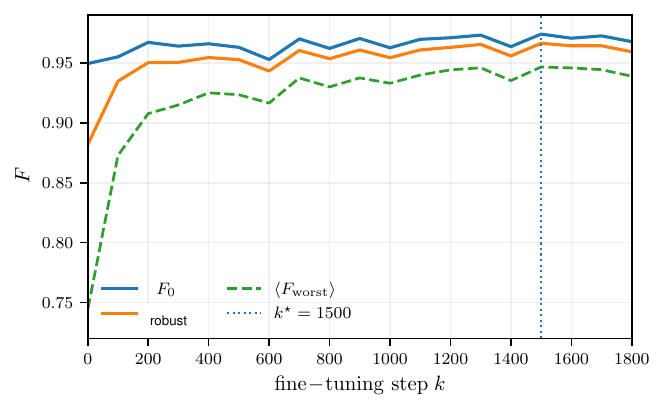}

\caption{\rev{Convergence of risk-aware fine-tuning.
The curves show nominal mean fidelity ($F_0$), mean
fidelity under the training uncertainty scenarios
($F_{\mathrm{robust}}$), and gate-averaged worst-scenario
fidelity ($\langle F_{\mathrm{worst}}\rangle$).
The selected checkpoint at step 1500 shows improved
robustness while retaining high nominal fidelity.}}

\label{fig:robust_convergence}
\end{figure}

Risk-aware fine-tuning was performed as a single fixed RF-gain training experiment with $|g-1|\le 0.05$. Fine-tuning uses AdamW with learning rate $1.5\times10^{-4}$, four gates per microbatch, four gradient-accumulation steps, and the same control regularizers as Eq.~\eqref{eq:nominal_loss}.
The exact fine-tuning and routine uncertainty settings are summarized in Table~\ref{tab:robust_hparams}. Figure~\ref{fig:robust_convergence} reports held-out convergence during the executed $\pm5\%$ risk-aware fine-tuning run: the robust validation mean rises from $0.8823$ at the first recorded step to a best value of $0.9664$ at step $1500$, while the nominal mean remains $0.9741$ at that selected checkpoint. The gate-averaged worst-scenario fidelity improves in parallel from 0.7451
to 0.9466, directly reflecting the adverse-scenario weighting introduced by the tail-risk term.  A complete derivation of the gradient used for this update is given in Appendix~\ref{app:backprop}.

\begin{table}[t]
\centering
\caption{Risk-aware fine-tuning and routine uncertainty configuration.  Quasi-static quantities are sampled once per pulse; only slice timing jitter varies within a pulse.}
\label{tab:robust_hparams}
\footnotesize
\renewcommand{\arraystretch}{1.08}
\setlength{\tabcolsep}{3.8pt}
\begin{tabular}{ll@{\hspace{0.8em}}ll}
\toprule
Training/risk & Value & Routine uncertainty & Range\\
\midrule
Initial LR & $1.5\times10^{-4}$ & RF gain $g-1$ & $\pm5\%$\\
Microbatch / accumulation & $4/4$ & Common $B_0$ & $\pm2\,\mathrm{Hz}$\\
Random / boundary scenarios & $8/2$ & Independent $\delta\nu_i$ & $\pm2\,\mathrm{Hz}$\\
CVaR tail mass $\alpha$ & $0.20$ & Independent $\delta J_{ij}$ & $\pm0.25\,\mathrm{Hz}$\\
$(\lambda_0,\lambda_\mu,\lambda_{\rm tail})$ & $(1,1,1.5)$ & Phase bias & $\pm1^\circ$\\
Maximum / selected step & $1800/1500$ & Clock error & $\pm2\,\mathrm{ppm}$\\
Training RF range & $\pm5\%$ only & Timing jitter & $10\,\mathrm{ps}$ RMS\\
Evaluation-only RF ranges & $\pm7.5,\pm10,\pm15\%$ & out of distribution (OOD) envelope & \\
\bottomrule
\end{tabular}
\end{table}

\section{Optimal-control baselines and benchmarking protocol}\label{sec:baselines}

To separate amortized compilation from per-target optimization, all baselines use the same Hamiltonian, $N=300$ I/Q slices, $35\,\mu\mathrm{s}$ step, radial amplitude map in Eq.~\eqref{eq:radial_bound}, fidelity in Eq.~\eqref{eq:unitary_fidelity}, and smoothness/power penalties.  The benchmark set contains 19 targets: 11 named gates and eight independently sampled Haar gates.

\paragraph{Nominal GRAPE.}
We optimize all $600$ pulse pixels directly for each target, using the same differentiable propagator as the network.  This is a GRAPE-style direct pixel optimization \citep{Khaneja2005OptimalControl}: the control variables are the time-slice amplitudes, while gradients are obtained by automatic differentiation through the matrix exponentials.  Six diversified nonzero initializations are optimized with Adam for at most $2000$ steps (initial learning rate $8\times10^{-4}$ with cosine decay), retaining the best iterate and stopping early when $F\ge0.9995$.  The best restart is optionally refined by L-BFGS for at most $200$ iterations.

\paragraph{Robust GRAPE.}
The robust baseline replaces nominal infidelity by the same composite objective as Eq.~\eqref{eq:robust_loss}, using the routine uncertainty envelope, eight random scenarios, and exact $\pm5\%$ RF boundaries.  It uses six restarts, up to $2200$ Adam steps at initial learning rate $6\times10^{-4}$, and the same L-BFGS refinement.  Candidate selection is performed on a common held-out set of 20 Sobol routine scenarios plus the two exact RF boundaries, avoiding selection on the restart-specific training scenarios.

\paragraph{Hybrid and secondary comparators.}
NN-seeded GRAPE starts from the nominal neural pulse and performs a short target-specific refinement (up to 400 Adam steps followed by at most 120 L-BFGS iterations). This quantifies how much per-target optimization remains after amortized compilation.

Timing is measured as wall-clock compilation time. For neural methods, the reported per-target latency is the online forward pass after training. For GRAPE-family methods the reported time includes all configured restarts and L-BFGS refinement.  \rev{The measured GRAPE latencies describe the specified multi-restart, high-fidelity optimization protocol; they should not be interpreted as the fastest possible GRAPE time to a lower target fidelity.}

NN seeding starts close to the high-fidelity region and reaches $F_{\rm gate}>0.999$ within roughly $120$ recorded iterations, while the random-start GRAPE procedures require substantially longer trajectories.  This convergence view complements the end-point benchmark in \rev{Section}~\ref{sec:results} and illustrates why a learned pulse can remain useful as an initializer even when final target-specific refinement is desired.

\section{Numerical results}\label{sec:results}

\subsection{Nominal generalization and uncertainty diagnosis}

\begin{figure}[t]
\centering
\begin{subfigure}[t]{0.485\linewidth}
  \centering\includegraphics[width=\linewidth]{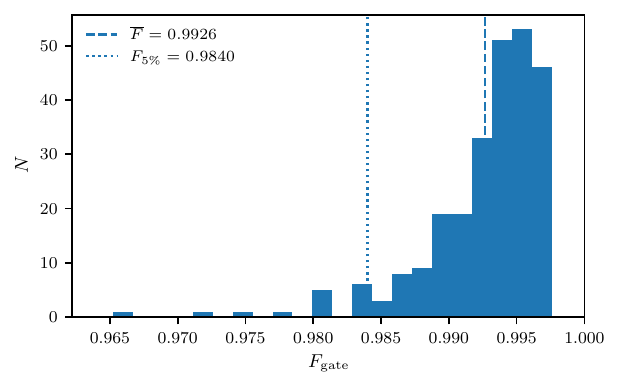}
  \caption{Held-out Haar targets.}
\end{subfigure}\hfill
\begin{subfigure}[t]{0.485\linewidth}
  \centering\includegraphics[width=\linewidth]{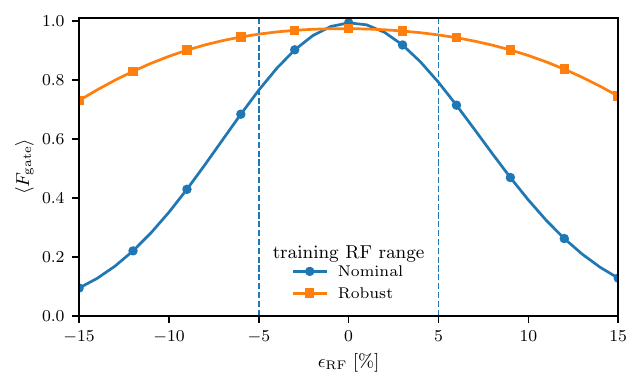}
  \caption{Pure RF-gain sweep.}
\end{subfigure}

\caption{\rev{Nominal accuracy and RF-gain robustness.
(a) Fidelity distribution over unseen Haar-distributed
targets, with mean and fifth percentile indicated.
(b) Mean fidelity versus RF-amplitude error for the
nominal and robust compilers. Dashed lines mark the
$\pm5\%$ training range. Robust fine-tuning broadens
the high-fidelity region beyond this range.}}

\label{fig:numerical_overview}
\end{figure}

The nominal checkpoint selected at step $48{,}600$ generalizes across the continuous target family.  On its fixed held-out Haar validation set it attains mean fidelity $\rev{0.9925}$, median $\rev{0.9941}$, and fifth percentile $\rev{0.9837}$; an independently generated Haar set gives a consistent mean of $\rev{0.9926}$.  Figure~\ref{fig:numerical_overview}(a) shows the latter distribution.  The stress test then separates nominal accuracy from sensitivity to model mismatch.  When several hardware uncertainties are applied simultaneously, the mean fidelity of the nominal compiler decreases from approximately 0.993 under ideal conditions to 0.920 under the routine uncertainty range and 0.577 under the more conservative stress-test range. Among the individual uncertainty sources examined, RF-amplitude errors produce the largest fidelity degradation, while chemical-shift, coupling, clock, and phase errors have smaller effects over their tested ranges.

\subsection{Risk-aware fine-tuning and out-of-range robustness}

Risk-aware fine-tuning exchanges a small amount of zero-error fidelity for a substantially broader uncertainty plateau.  On the final full-Haar evaluation suite, nominal-condition mean fidelity changes from $\rev{0.9930}$ for the original compiler to $\rev{0.9736}$ for the robust compiler.  Under the routine joint $\pm5\%$ RF suite, however, the mean increases from $\rev{0.9190}$ to $\rev{0.9660}$ and the gate-averaged worst-scenario mean from $\rev{0.7761}$ to $\rev{0.9422}$. Because the selected model was trained only within the $\pm5\%$ RF envelope, the wider suites test extrapolation: at $\pm7.5\%$ and $\pm10\%$ RF the robust means are $\rev{0.9572}$ and $\rev{0.9439}$, compared with $\rev{0.8413}$ and $\rev{0.7512}$ for the nominal model. Under the conservative $\pm15\%$ out of distribution (OOD) suite the corresponding means are $\rev{0.8951}$ and $\rev{0.5750}$.  The pure RF sweep in \rev{Figure}~\ref{fig:numerical_overview}(b) shows the same effect without the other uncertainty channels: \rev{the robust mean is $0.8796$ at $-10\%$ and $0.8822$ at $+10\%$, whereas the nominal mean is $0.3510$ and $0.3925$ at the same respective settings.}

\subsection{Comparison with per-target optimal control}
\begin{figure}[t]
\centering
\begin{subfigure}[t]{0.485\linewidth}\centering
\includegraphics[width=\linewidth]{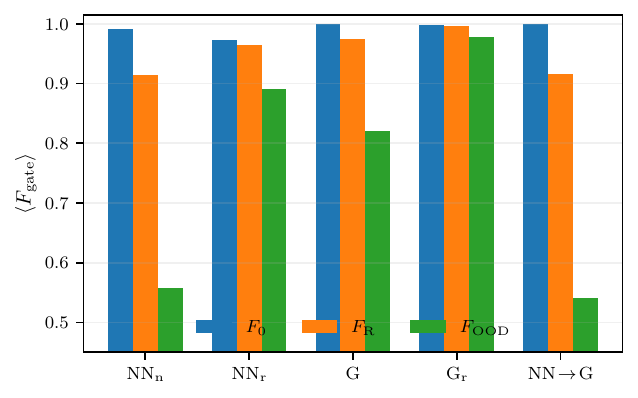}
\caption{Nominal, Robust, and OOD fidelity.}
\end{subfigure}\hfill
\begin{subfigure}[t]{0.485\linewidth}\centering
\includegraphics[width=\linewidth]{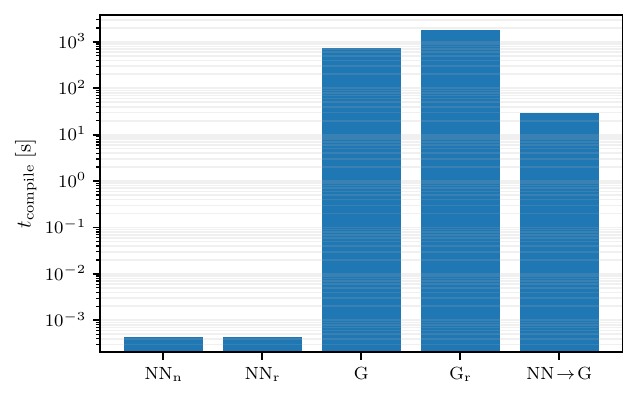}
\caption{Per-target compilation latency (log scale).}
\end{subfigure}

\caption{\rev{Comparison of neural compilation and
GRAPE-based optimization on 19 target gates.
(a) Mean fidelity under nominal ($F_0$), routine
uncertainty ($F_R$), and conservative out-of-distribution
($F_{\mathrm{OOD}}$) conditions.
(b) Per-target compilation time on a logarithmic scale.
Neural-network times exclude offline training, while
GRAPE times include target-specific optimization.
Robust GRAPE provides the highest uncertainty-tested
fidelity, whereas neural inference offers substantially
lower online latency.}}

\label{fig:benchmark}
\end{figure}

\rev{Figure}~\ref{fig:benchmark} summarizes the matched 19-gate benchmark.  Strong nominal GRAPE obtains the highest nominal mean among non-robust methods, $\rev{0.9995}$, and robust GRAPE is the strongest method under the uncertainty suites, reaching $\rev{0.9973}$ in the routine joint test and $\rev{0.9777}$ in the conservative OOD test. These results separate the two operating regimes: target-specific GRAPE
provides the highest offline fidelity, while the learned compiler provides amortized waveform
generation for new targets. Its advantage is amortization: the nominal and robust networks require approximately $0.44\,\mathrm{ms}$ for one forward compilation in this implementation, \rev{whereas the configured six-restart GRAPE procedures, targeting fidelity $F\ge 0.9995$ for the nominal baseline, require hundreds to thousands of seconds per target under the reported benchmark conditions.}  NN-seeded GRAPE reaches mean fidelity $\rev{0.9994}$ in about $30\,\mathrm{s}$, showing that a learned pulse is also a useful warm start; because that refinement is nominal, its robustness reverts close to the nominal network. 

\section{Experimental NMR validation}
\label{sec:experiment}

We next test experimentally whether the proposed neural compiler can map
continuously parameterized single-qubit gates to executable NMR control
waveforms. Experiments were carried out on a three-spin benchtop NMR
processor based on a $C_2F_3I$ molecule, in which the three $^{19}$F nuclear
spins serve as qubits. In NMR, the transverse magnetization of the spin
ensemble is detected in the time domain as a free-induction decay (FID).
Fourier transformation converts the FID into a frequency-domain spectrum.
The resonance frequencies are primarily determined by the chemical shifts,
while the fine splittings arise from scalar $J$ couplings between the
spins~\citep{Vandersypen2005NMRTechniques}. The Larmor frequency of the nuclear spins in this device is approximately 26 MHz. Accordingly, the frequency axes of all experimental spectra are plotted after subtracting this common 26 MHz carrier frequency. This convention is consistent with the rotating-frame Hamiltonian used throughout the analysis, in which the large common Larmor frequency is removed by the rotating-frame transformation and only the relative frequency offsets remain.

\begin{figure}[t]
    \centering
    \includegraphics[width=0.8\linewidth]{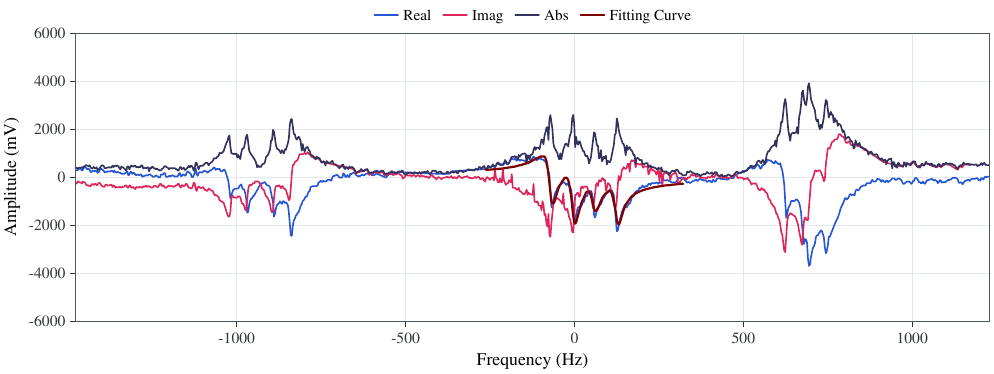}
    \caption{
    Frequency-domain spectrum of the three-spin $^{19}$F NMR system after a
    short excitation pulse. Each of the three resonance groups corresponds
    primarily to one fluorine spin, and scalar coupling to the other two
    spins splits each resonance into four single-quantum transitions. The
    resulting twelve transition channels form the spectral observables used
    in the tomography reconstruction.
    }
    \label{12peaks}
\end{figure}

Each spectral line can be
associated with a single-quantum transition between two spin states. A
single-quantum transition flips one observed spin while leaving the states
of the other spins unchanged. Since the present system contains three
spin-$1/2$ nuclei, each observed spin has four possible configurations of
the other two spins and therefore contributes four allowed transitions.
For example, the four transitions associated with spin~1 are
\begin{equation}
|000\rangle\leftrightarrow|100\rangle,\qquad
|001\rangle\leftrightarrow|101\rangle,\qquad
|010\rangle\leftrightarrow|110\rangle,\qquad
|011\rangle\leftrightarrow|111\rangle.
\end{equation}
Analogous groups of four transitions occur for spins 2 and 3, giving twelve
single-quantum transitions in total, as shown in \rev{Figure}~\ref{12peaks}. Their frequency separation is produced
by the chemical shifts and scalar couplings, while their complex amplitudes
contain the information used for quantum-state reconstruction.

The platform prepares a pseudo-pure state (PPS) using its built-in
calibration and control routines~\citep{Spinq}. In liquid-state NMR, the
ensemble density matrix can be written in the form
\begin{equation}
\rho_{\mathrm{NMR}}
=
\frac{I_8}{8}
+
\epsilon\,\widetilde{\rho},
\label{eq:nmr_deviation}
\end{equation}
where $\epsilon\ll1$ is the ensemble polarization and
$\widetilde{\rho}$ is the traceless deviation density matrix. The identity
component does not contribute to the detected NMR signal. Accordingly, throughout the experimental analysis we reconstruct and
compare only the deviation density matrix $\widetilde{\rho}$. For the ideal PPS,
\begin{equation}
\widetilde{\rho}_{\mathrm{PPS}}^{\,\mathrm{id}}
=
|000\rangle\langle000|
-
\frac{I_8}{8}.
\label{eq:pps_deviation}
\end{equation}

We first test the continuously parameterized family
\begin{equation}
\label{Param}
U_2(\gamma,\theta,\alpha)
=
\cos\!\frac{\gamma}{2}\,I_2
-i\sin\!\frac{\gamma}{2}
\left(
n_x\sigma_x+n_y\sigma_y+n_z\sigma_z
\right),
\qquad
\boldsymbol{n}
=
(\sin\theta\cos\alpha,
 \sin\theta\sin\alpha,
 \cos\theta),
\end{equation}
by fixing $\theta=90^\circ$ and $\gamma=90^\circ$ while varying $\alpha$.
For each value of $\alpha$, the corresponding neural-network-generated
waveform is applied to the experimentally prepared PPS.

\begin{figure*}[t]
    \centering
    \begin{subfigure}{0.485\linewidth}
        \includegraphics[width=\linewidth]{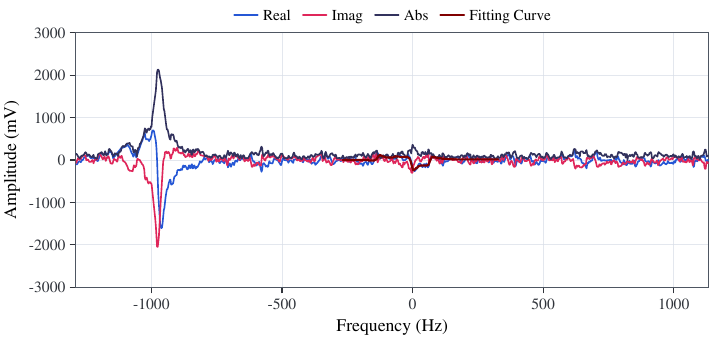}
        \caption{$\alpha=20^\circ$}
    \end{subfigure}
    \hfill
    \begin{subfigure}{0.485\linewidth}
        \includegraphics[width=\linewidth]{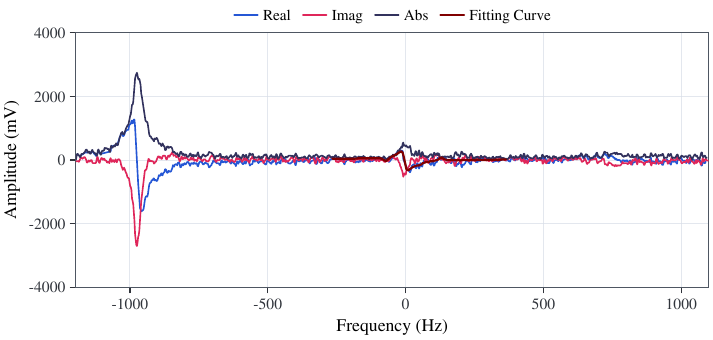}
        \caption{$\alpha=40^\circ$}
    \end{subfigure}

    \begin{subfigure}{0.485\linewidth}
        \includegraphics[width=\linewidth]{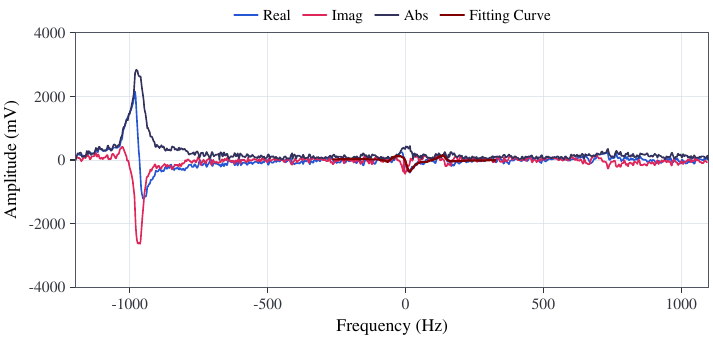}
        \caption{$\alpha=60^\circ$}
    \end{subfigure}
    \hfill
    \begin{subfigure}{0.485\linewidth}
        \includegraphics[width=\linewidth]{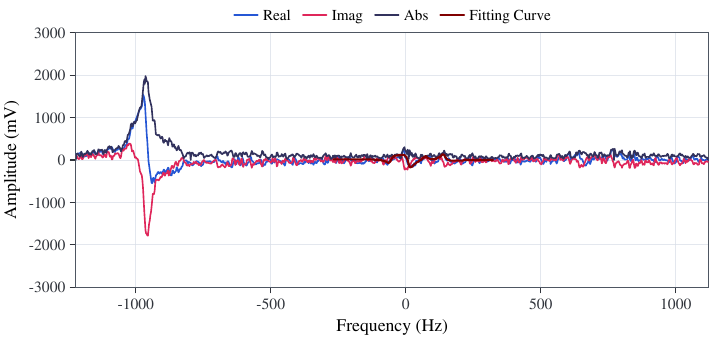}
        \caption{$\alpha=80^\circ$}
    \end{subfigure}

    \caption{
    Experimental frequency-domain NMR spectra obtained after applying the
    neural-network-generated single-qubit gate to the PPS for several values
    of $\alpha$ for the gate family in Eq.~\eqref{Param}. The parameters $\theta=90^\circ$ and $\gamma=90^\circ$
    are fixed while the rotation-axis azimuth $\alpha$ is varied. The real
    and imaginary traces represent the two quadratures of the detected
    complex NMR signal.
    }
    \label{alpha_sweep}
\end{figure*}

\begin{figure}[t]
    \centering
    \includegraphics[width=0.45\linewidth]{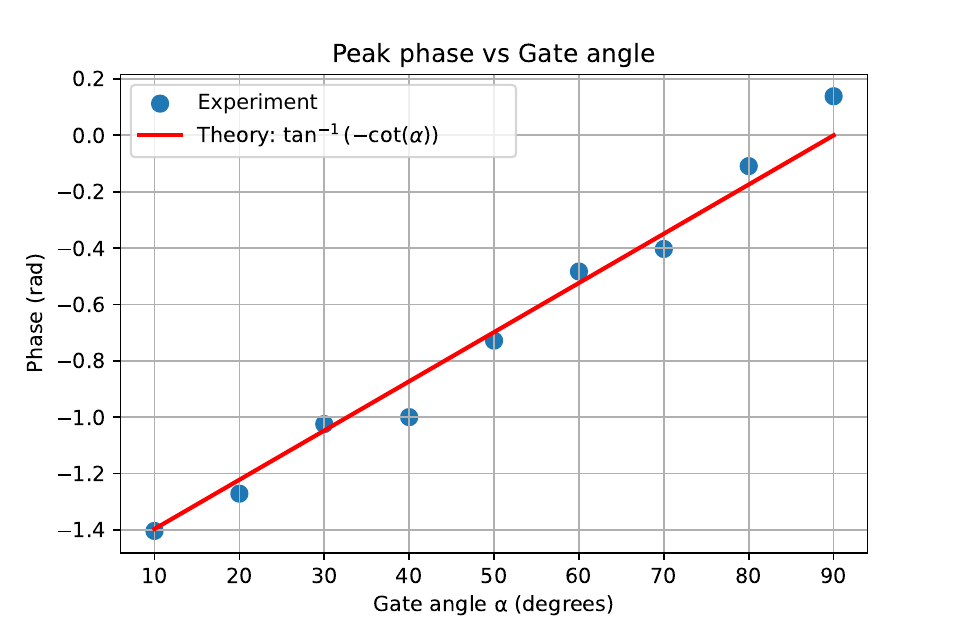}
    \caption{
    Experimental phase of the integrated complex NMR signal as a function of
    $\alpha$ for the gate family in Eq.~\eqref{Param}. The phase is
    evaluated from the real and imaginary quadratures using
    Eq.~\eqref{eq:spectral_phase}. The corresponding analytical dependence
    is derived by~\citet{PrevWork}. Using $\alpha$ in radians, the experimental phase dependence is well described by the linear fit
$
\phi=(1.09\pm0.05)\alpha-(1.65\pm0.05), R^2\simeq0.985,
$
which is consistent with the theoretical relation
$
\phi=\tan^{-1}[-\cot(\alpha)]=\alpha-\frac{\pi}{2}
$.
    }
    \label{PhaseOut}
\end{figure}

For this test, the complex amplitudes of the transitions associated with the
addressed spin are obtained by integrating fixed frequency windows in the
real and imaginary spectra. If an integrated complex signal is written as
\begin{equation}
A=A_{\mathrm{R}}+iA_{\mathrm{I}},
\end{equation}
its experimental phase is evaluated as
\begin{equation}
\phi_{\mathrm{spec}}
=
\operatorname{atan2}
\left(
A_{\mathrm{I}},A_{\mathrm{R}}
\right).
\label{eq:spectral_phase}
\end{equation}
The use of $\operatorname{atan2}$ retains the correct phase quadrant. As
shown analytically by ~\citet{PrevWork}, the spectral phase varies linearly with the azimuthal control parameter $\alpha$ for the family
in Eq.~\eqref{Param}. The experimental spectra in
\rev{Figure}~\ref{alpha_sweep} and the extracted phases in \rev{Figure}~\ref{PhaseOut}
confirm the predicted dependence.

\rev{Although the conventional Hadamard matrix has determinant $-1$, its phase-equivalent representative $iH$ belongs to $\SU(2)$ and implements the same physical operation up to global phase.}
The final hardware test considers an in-distribution Hadamard target acting
on spin~1,
\begin{equation}
H_1=H\otimes I\otimes I,
\qquad
|\psi_H\rangle
=
|+00\rangle
=
\frac{|000\rangle+|100\rangle}{\sqrt{2}},
\label{eq:had_target}
\end{equation}
whose ideal deviation density matrix is
\begin{equation}
\widetilde{\rho}_{H}^{\,\mathrm{id}}
=
|\psi_H\rangle\langle\psi_H|
-
\frac{I_8}{8}.
\label{eq:had_deviation}
\end{equation}

\begin{figure}[t]
\centering
\begin{subfigure}[t]{0.49\linewidth}
    \centering
    \includegraphics[width=\linewidth]{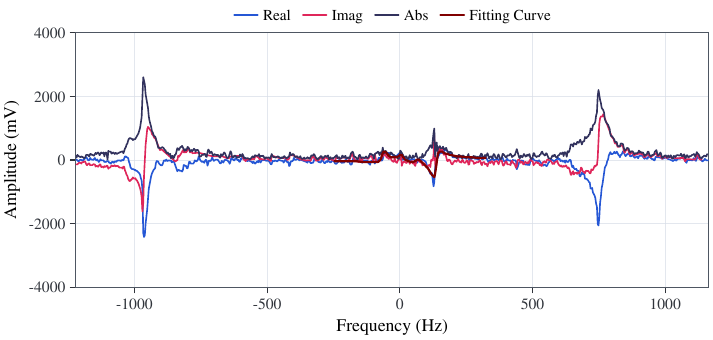}
    \caption{Nominal +10\%.}
\end{subfigure}
\hfill
\begin{subfigure}[t]{0.49\linewidth}
    \centering
    \includegraphics[width=\linewidth]{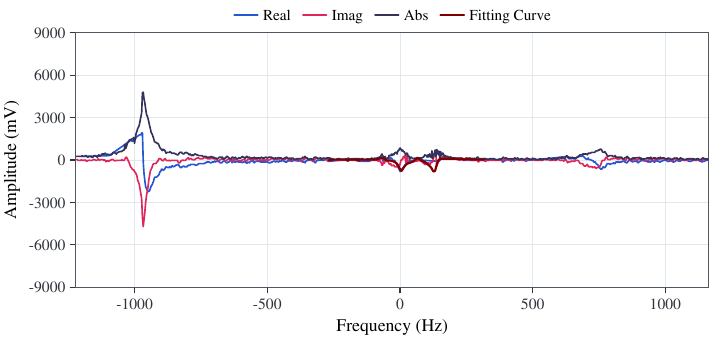}
    \caption{Robust +10\%.}
\end{subfigure}
\caption{
Comparison of the readout spectra at +10\% commanded RF-amplitude error obtained after applying the Hadamard gate to the PPS. The nominal pulse produces pronounced responses in additional resonance regions, indicating stronger off-target spin excitation, whereas the robust pulse preserves a cleaner response associated with the intended operation.
}
\label{fig:III_NP10_RP10}
\end{figure}

For each network, the two predicted RF quadratures are converted to amplitude
and phase according to
\begin{equation}
a_k
=
\sqrt{u_{x,k}^{\,2}+u_{y,k}^{\,2}},
\qquad
\varphi_k
=
\operatorname{atan2}(u_{y,k},u_{x,k}).
\end{equation}
The commanded global RF-amplitude error is incorporated into the waveform as
\begin{equation}
a_k
\longmapsto
(1+\epsilon_{\mathrm{RF}})a_k,
\end{equation}
and measurements are acquired for
\begin{equation}
\epsilon_{\mathrm{RF}}
\in
\{-10,-5,0,+5,+10\}\%
\end{equation}
for both the nominal and robust neural pulses. A direct comparison of the readout spectra at $+10\%$ RF-amplitude error further illustrates the effect of robust training, \rev{Figure}~\ref{fig:III_NP10_RP10}. For the nominal pulse, appreciable spectral responses appear around the intended transitions and in additional resonance groups, indicating unwanted excitation of additional spin modes. In contrast, the robust pulse strongly suppresses these unintended responses and concentrates the spectral weight predominantly in the expected spin-1 resonance region. This reduced off-target excitation is consistent with the larger deviation-matrix correlation obtained for the robust pulse at $+10\%$ RF error.

\subsection{Deviation-density-matrix tomography}

Full three-qubit tomography is performed after each experimental pulse.
Eleven pre-readout operations,
\begin{equation}
\{
III,\,
XII,\,
IXI,\,
IIX,\,
IXX,\,
XXX,\,
YII,\,
IYI,\,
IIY,\,
YYI,\,
YYY
\},
\end{equation}
interrogate the twelve complex single-quantum transitions. Here X and Y denote 90◦ pre-readout rotations about the corresponding
axes, while III denotes no pre-readout rotation. Since both real
and imaginary quadratures are retained, the complete measurement vector
contains
\begin{equation}
11\times12\times2=264
\end{equation}
real values.

Let $\{B_\mu\}_{\mu=1}^{63}$ be a Hermitian traceless basis for the
three-qubit operator space. The reconstructed experimental deviation matrix
is expanded as
\begin{equation}
\widetilde{\rho}_{\mathrm{exp}}
=
\sum_{\mu=1}^{63}r_\mu B_\mu.
\label{eq:deviation_expansion}
\end{equation}
For each pre-readout rotation, the real and imaginary components of the
twelve observed transition amplitudes are appended to the measurement vector
$\mathbf f$. The tomography problem can then be written as
\begin{equation}
\mathbf f
\approx
M\mathbf r,
\label{eq:tomo_linear}
\end{equation}
where
$\mathbf r=(r_1,\ldots,r_{63})^{T}$ and $M$ is constructed from the known
pre-readout rotations, the Hermitian basis, and the twelve observed
single-quantum transition matrix elements. The coefficients are obtained by
linear least squares,
\begin{equation}
\widehat{\mathbf r}
=
\operatorname*{arg\,min}_{\mathbf r}
\|M\mathbf r-\mathbf f\|_2^2,
\label{eq:tomo_lstsq}
\end{equation}
after which Eq.~\eqref{eq:deviation_expansion} gives
$\widetilde{\rho}_{\mathrm{exp}}$.

The same spectral-processing procedure is used for the PPS reference and
for every nominal, robust, and RF-perturbed experiment. Each transition
amplitude is obtained by integrating a fixed frequency window independently
in the real and imaginary quadratures. Receiver phase is calibrated once from the PPS reference and then kept fixed
for all subsequent experimental conditions. Because the identity component of Eq.~\eqref{eq:nmr_deviation} is
unobservable, experiment and theory are compared directly in deviation-matrix
space. We use the normalized Hilbert--Schmidt correlation
\begin{equation}
C_{\mathrm{HS}}
\left(
\widetilde{\rho}_1,\widetilde{\rho}_2
\right)
=
\frac{
\operatorname{Re}
\left[
\operatorname{Tr}
\left(
\widetilde{\rho}_1\widetilde{\rho}_2
\right)
\right]
}{
\sqrt{
\operatorname{Tr}
\left(
\widetilde{\rho}_1^2
\right)
\operatorname{Tr}
\left(
\widetilde{\rho}_2^2
\right)
}
}.
\label{eq:HS_corr}
\end{equation}
For Hermitian deviation matrices,
\begin{equation}
-1\le C_{\mathrm{HS}}\le1,
\end{equation}
with $C_{\mathrm{HS}}=1$ corresponding to identical deviation-matrix
structure up to an overall positive scale. The normalization therefore makes
the comparison insensitive to a common multiplicative signal factor, as is
appropriate for the present NMR measurements. As the reconstructed object is the traceless deviation matrix $\tilde{\rho}$, rather than the normalized full ensemble density operator $\rho_{\mathrm{NMR}}$, it satisfies $\operatorname{Tr}(\tilde{\rho}) = 0$ by construction. Its eigenvalues and $\operatorname{Tr}(\tilde{\rho}^{\,2})$ therefore characterize the reconstructed deviation signal and should not be interpreted as the eigenvalues or purity of a unit-trace physical density operator. For this reason, we quantify agreement using the normalized Hilbert--Schmidt correlation in Eq.~\eqref{eq:HS_corr}, which is defined directly in deviation-matrix space.

For the experimentally prepared PPS we define
\begin{equation}
C_{\mathrm{PPS}}
=
C_{\mathrm{HS}}
\left(
\widetilde{\rho}_{\mathrm{PPS}}^{\,\mathrm{exp}},
\widetilde{\rho}_{\mathrm{PPS}}^{\,\mathrm{id}}
\right),
\label{eq:Cpps}
\end{equation}
and obtain
\begin{equation}
C_{\mathrm{PPS}}=0.82774.
\label{eq:Cpps_value}
\end{equation}

For each Hadamard experiment we analogously report
\begin{equation}
C_H(\epsilon_{\mathrm{RF}})
=
C_{\mathrm{HS}}
\left(
\widetilde{\rho}_{H}^{\,\mathrm{exp}}
(\epsilon_{\mathrm{RF}}),
\widetilde{\rho}_{H}^{\,\mathrm{id}}
\right).
\label{eq:CH}
\end{equation}

To compare the numerical simulations with experiment using the same metric,
the simulated density matrix is first expressed in deviation form,
\begin{equation}
\widetilde{\rho}_{H}^{\,\mathrm{sim}}
=
\rho_{H}^{\,\mathrm{sim}}
-
\frac{I_8}{8}.
\end{equation}
\rev{For the ideal Hadamard output $|\psi_H\rangle$ and simulated output state $\rho_H^{\mathrm{sim}}$, define $F_{\mathrm{state}}=\langle\psi_H|\rho_H^{\mathrm{sim}}|\psi_H\rangle$. For the pure simulated output states considered here, the normalized Hilbert--Schmidt correlation is then given by}
\begin{equation}
C_{\mathrm{HS}}^{\mathrm{sim}}
=
\frac{8F_{\mathrm{state}}-1}{7}.
\label{eq:sim_hs_conversion}
\end{equation}
 The simulation and tomography values in Table~\ref{tab:hardware_results}
are therefore expressed in the same deviation-matrix metric.

\begin{table}[t]
\centering
\caption{
Normalized Hilbert--Schmidt correlation between the simulated or
experimentally reconstructed Hadamard-output deviation matrix and the ideal
target $\widetilde{\rho}_{H}^{\,\mathrm{id}}$. Experimental values are
obtained directly from the linear reconstruction of the NMR deviation
density matrix.
}
\label{tab:hardware_results}
\footnotesize
\renewcommand{\arraystretch}{1.08}
\begin{tabular}{rcccc}
\toprule
$\epsilon_{\mathrm{RF}}$
& Sim. nominal
& Exp. nominal
& Sim. robust
& Exp. robust\\
\midrule
$-10\%$ & 0.40274 & 0.40246 & 0.95423 & 0.77406\\
$-5\%$  & 0.79739 & 0.70080 & 0.98313 & 0.75934\\
$0\%$   & 0.99006 & 0.78325 & 0.97910 & 0.82022\\
$+5\%$  & 0.80069 & 0.63902 & 0.97151 & 0.79720\\
$+10\%$ & 0.39151 & 0.36937 & 0.95022 & 0.78067\\
\bottomrule
\end{tabular}
\end{table}

\begin{figure}[t]
\centering
\includegraphics[width=0.5\linewidth]
{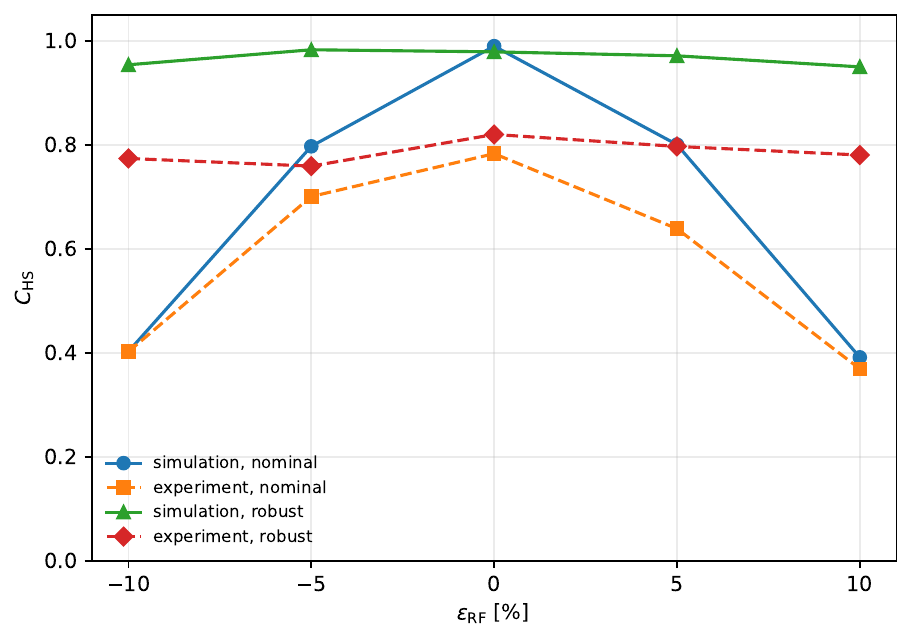}
\caption{
Normalized Hilbert--Schmidt correlation with the ideal Hadamard deviation
matrix as a function of commanded RF-amplitude error. Simulated and
experimental results are expressed using the same deviation-matrix
correlation defined in Eq.~\eqref{eq:HS_corr}. The nominal control exhibits
a narrow response around the calibrated RF amplitude, while the robust
control maintains a broad high-correlation plateau across the full
$\pm10\%$ experimental sweep.
}
\label{fig:hardware}
\end{figure}

\rev{Figure~\ref{fig:hardware} compares the simulated and
experimentally reconstructed Hadamard-output deviation-matrix
correlations across the RF-amplitude sweep, showing the narrow
response of the nominal pulse and the broader RF-gain tolerance
of the robust pulse.}
 The nominal pulse is strongly localized around the calibrated
RF amplitude: its experimental correlation reaches
\begin{equation}
C_H^{\mathrm{nom}}(0)=0.78325
\end{equation}
at zero commanded error and decreases to
\begin{equation}
C_H^{\mathrm{nom}}(-10\%)=0.40246,
\qquad
C_H^{\mathrm{nom}}(+10\%)=0.36937
\end{equation}
at the two endpoints. In contrast, the robust pulse maintains a substantially
flatter experimental response,
\begin{equation}
C_H^{\mathrm{rob}}(-10\%)=0.77406,
\qquad
C_H^{\mathrm{rob}}(0)=0.82022,
\qquad
C_H^{\mathrm{rob}}(+10\%)=0.78067.
\end{equation}
At $-10\%$ RF error, robustification therefore increases the experimentally
reconstructed correlation by
\begin{equation}
0.77406-0.40246=0.37160.
\end{equation}
At $+10\%$, the corresponding increase is
\begin{equation}
0.78067-0.36937=0.41130.
\end{equation}

\begin{figure}[t]
\centering
\begin{subfigure}[t]{0.49\linewidth}
\centering
\includegraphics[width=\linewidth]{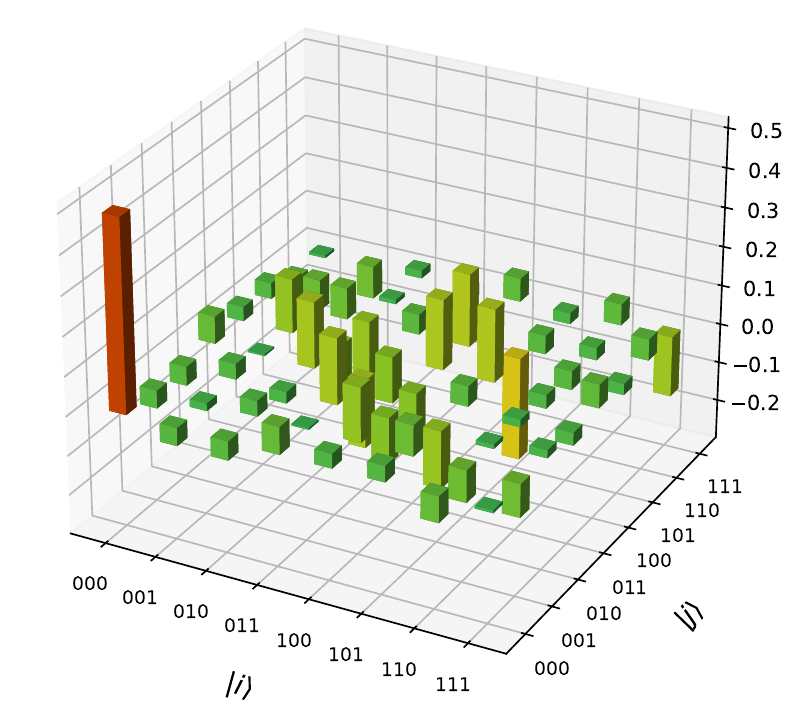}
\caption{
Nominal, $+10\%$: $C_H=0.3694$.
}
\end{subfigure}
\hfill
\begin{subfigure}[t]{0.49\linewidth}
\centering
\includegraphics[width=\linewidth]{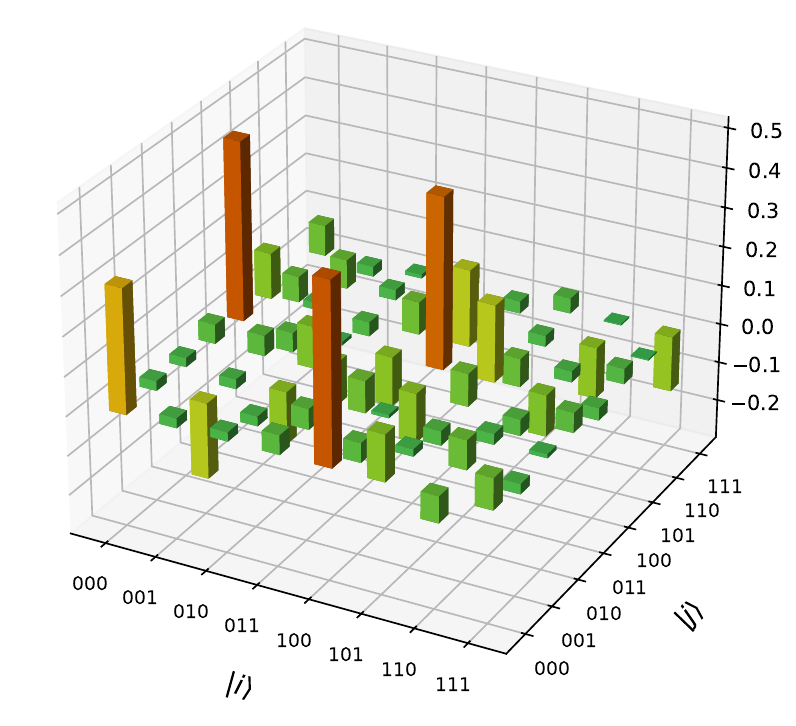}
\caption{
Robust, $+10\%$: $C_H=0.7807$.
}
\end{subfigure}
\caption{
Real parts of the reconstructed experimental deviation density matrices at
$+10\%$ commanded RF-amplitude error. The robust pulse preserves the
characteristic Hadamard deviation-matrix structure substantially more
strongly than the nominal pulse under the imposed RF-amplitude perturbation.
}
\label{fig:density}
\end{figure}
Figure~\ref{fig:density} provides a complementary matrix-level comparison at
the largest positive RF perturbation. For the ideal Hadamard output, the
deviation matrix contains characteristic contributions associated with the
$|000\rangle$ and $|100\rangle$ populations and their
$|000\rangle\!\leftrightarrow|100\rangle$ coherence. 

The hardware experiments probe two aspects of the neural compiler. Across the continuously parameterized $\alpha$ sweep, the network-generated controls reproduce the expected systematic spectral-phase dependence. The Hadamard RF-amplitude sweep tests the robustness prediction that motivated the risk-aware training: the nominal correlation decreases sharply away from calibration, while the robust correlation remains comparatively flat across the measured range. The observed plateau is consistent with the RF-gain tolerance predicted by the numerical model after transfer of the learned pulse sequence to the three-spin NMR processor. The experimental tomography reported here characterizes output deviation density matrices for selected input states and therefore does not constitute full quantum process tomography of the implemented gate. Accordingly, the experimental Hilbert–Schmidt
correlations should not be interpreted as process fidelities. The measurements test the two hardware predictions central to the present study: the systematic parameter dependence of the compiled continuous gate family and the broadening of the RF-gain tolerance produced by robust training. Full process tomography over representative compiled gates would provide a complementary characterization of the implemented quantum channel.

\subsection{Reconstruction uncertainty}

\begin{table*}[t]
\centering
\caption{
Experimental Hilbert--Schmidt correlations with Monte Carlo uncertainty,
systematic-sensitivity diagnostics, and combined effective uncertainty.
}
\label{tab:experimental_uncertainty_combined}
\footnotesize
\renewcommand{\arraystretch}{1.06}
\setlength{\tabcolsep}{1.0pt}
\resizebox{\textwidth}{!}{%
\begin{tabular}{lccccccc}
\toprule
Condition
& $C_{\mathrm{HS}}$
& MC std.
& MC 95\% interval
& Routine systematic range
& Spin 3 overlap
& Overall sensitivity envelope
& Effective SD / joint 95\% interval \\
\midrule

PPS
& 0.82774
& 0.00304
& [0.82077,\;0.83136]
& [0.81818,\;0.82987]
& [0.82774,\;0.89728]
& [0.81818,\;0.89728]
& 0.01450 / [0.82378,\;0.88055] \\

\midrule

Nominal $-10\%$
& 0.40246
& 0.01512
& [0.37065,\;0.42847]
& [0.33318,\;0.40866]
& [0.36397,\;0.41303]
& [0.33318,\;0.42847]
& 0.02470 / [0.32423,\;0.41910] \\

Nominal $-5\%$
& 0.70080
& 0.00752
& [0.68772,\;0.71343]
& [0.65827,\;0.70499]
& [0.66876,\;0.72123]
& [0.65827,\;0.72123]
& 0.01634 / [0.65872,\;0.72128] \\

Nominal $0\%$
& 0.78325
& 0.00367
& [0.77395,\;0.78674]
& [0.76625,\;0.78677]
& [0.76169,\;0.80402]
& [0.76169,\;0.80402]
& 0.01065 / [0.77126,\;0.81320] \\

Nominal $+5\%$
& 0.63902
& 0.00653
& [0.62679,\;0.64844]
& [0.62972,\;0.65453]
& [0.63902,\;0.66478]
& [0.62679,\;0.66478]
& 0.01321 / [0.62719,\;0.67757] \\

Nominal $+10\%$
& 0.36937
& 0.00653
& [0.35839,\;0.38079]
& [0.36038,\;0.39077]
& [0.35137,\;0.39024]
& [0.35137,\;0.39077]
& 0.01337 / [0.34741,\;0.39901] \\

\midrule

Robust $-10\%$
& 0.77406
& 0.00557
& [0.76059,\;0.78179]
& [0.74442,\;0.77816]
& [0.76558,\;0.80887]
& [0.74442,\;0.80887]
& 0.01389 / [0.74793,\;0.80197] \\

Robust $-5\%$
& 0.75934
& 0.00475
& [0.74939,\;0.76719]
& [0.74213,\;0.76250]
& [0.72790,\;0.79437]
& [0.72790,\;0.79437]
& 0.01228 / [0.74069,\;0.78862] \\

Robust $0\%$
& 0.82022
& 0.00228
& [0.81384,\;0.82210]
& [0.80612,\;0.82378]
& [0.80022,\;0.83686]
& [0.80022,\;0.83686]
& 0.00795 / [0.80872,\;0.84023] \\

Robust $+5\%$
& 0.79720
& 0.00307
& [0.79028,\;0.80062]
& [0.77756,\;0.80098]
& [0.75922,\;0.81373]
& [0.75922,\;0.81373]
& 0.00902 / [0.77959,\;0.81485] \\

Robust $+10\%$
& 0.78067
& 0.00206
& [0.77455,\;0.78209]
& [0.75935,\;0.78518]
& [0.76326,\;0.80492]
& [0.75935,\;0.80492]
& 0.01020 / [0.76360,\;0.80417] \\

\bottomrule
\end{tabular}%
}
\end{table*}

The uncertainty of the experimentally reconstructed states was analyzed at two complementary levels: first, by propagating plausible shot-to-shot spectral fluctuations through the complete tomography procedure, and second, by assessing the sensitivity of the reconstruction to systematic features of the spectral analysis. These contributions were subsequently incorporated into a joint Monte Carlo analysis to obtain a single effective reconstruction-sensitivity interval for each experimental condition \citep{Khirich2021MonteCarlo}.

Experimental reconstruction uncertainty was first estimated by Monte Carlo propagation through the complete tomography pipeline \citep{Khirich2021MonteCarlo}. For each realization, every digitized real and imaginary spectral value was perturbed by an independent bounded relative variation,
\begin{equation}
S_i^{(m)}
=
S_i\left(1+\delta_i^{(m)}\right),
\qquad
\delta_i^{(m)}
\sim
\mathcal{U}(-0.02,0.02),
\end{equation}
corresponding to a conservative $\pm2\%$ spectral-amplitude uncertainty. In addition, a slowly varying frequency-position uncertainty was included to account for possible spectral drift during the sequence of tomography acquisitions. The frequency offsets at the first and last of the eleven readout experiments were independently sampled from
\begin{equation}
\Delta f_{\mathrm{start}},
\Delta f_{\mathrm{end}}
\sim
\mathcal{U}(-5,5)\ {\rm Hz},
\end{equation}
and the offsets for the intermediate readouts were obtained by linear interpolation across the eleven experiments. The same frequency offset was applied to the paired real and imaginary quadratures of a given readout.

Each perturbed dataset was then processed using exactly the same fixed integration windows, receiver-phase corrections, global signal scaling, and linear deviation-density-matrix reconstruction used for the nominal experimental data \citep{Leskowitz2004StateInterrogation,Zhang2012FrustratedMagnet}. A total of $10^4$ Monte Carlo reconstructions was generated for each experimental condition, from which the corresponding distribution of the normalized Hilbert--Schmidt correlation $C_{\mathrm{HS}}$ was evaluated. During repeated observations of the NMR signal under otherwise unchanged experimental conditions, no substantial spectral-amplitude fluctuations or frequency displacements were observed. Since performing and storing a sufficiently large number of complete repeated tomography acquisitions would be experimentally demanding, the reconstruction uncertainty was therefore estimated numerically by perturbing the measured spectra within conservative amplitude and frequency ranges and propagating these perturbations through the complete tomography procedure \citep{Khirich2021MonteCarlo}. 

 The lower absolute experimental correlations should also be interpreted in the context of limitations of the present NMR measurement and reconstruction \citep{Zhang2012FrustratedMagnet}. In particular, some transitions, especially the central peaks associated with spin~3, are only partially resolved and consequently contain overlapping spectral contributions that cannot be perfectly separated by fixed-window integration \citep{Zhang2012FrustratedMagnet}. Such overlap can introduce a persistent reconstruction bias. To examine these effects separately from the stochastic spectral fluctuations, systematic sensitivity was assessed by modifying individual elements of the analysis and repeating the reconstruction. The tested variations included alternative baseline treatments, receiver-phase perturbations, shifts of the integration-window positions, and the influence of the partially mixed central spin-3 transitions. The receiver phases were varied within $\pm1^\circ$, while the integration-window positions were varied by approximately one digitized frequency bin, corresponding to $\pm2.56$~Hz. The influence of the overlapping central spin-3 transitions was additionally examined through conservative reconstruction diagnostics in which their contribution was partially or completely reduced. These tests quantify the sensitivity of $C_{\mathrm{HS}}$ to plausible systematic reconstruction choices. Table~\ref{tab:experimental_uncertainty_combined} therefore reports both the stochastic Monte Carlo interval and the corresponding systematic-sensitivity ranges, together with a conservative envelope spanning the tested reconstruction variations.

Finally, to obtain a single effective uncertainty for each reconstructed experimental correlation, all identified fluctuation and reconstruction-sensitivity sources were propagated jointly through the tomography procedure using $10^4$ Monte Carlo realizations \citep{Khirich2021MonteCarlo}. The joint model simultaneously included all perturbations described above. The experimentally reconstructed values of $C_{\mathrm{HS}}$ were retained as the central values, while the $2.5$--$97.5$ percentile range of the resulting joint distribution was taken as the effective uncertainty interval, last column in \rev{Table}~\ref{tab:experimental_uncertainty_combined}. Across the joint reconstruction model, the robust and nominal correlation
intervals remain clearly separated at the $\pm 10\%$ RF endpoints, so the principal robustness comparison is preserved under the tested reconstruction variations. In the Monte Carlo analysis, the identified uncertainty sources were propagated at the level of the integrated tomography observables rather than by explicitly generating and reintegrating a complete perturbed spectrum for every realization. Because spectral integration is a linear operation, the mean and covariance of the twelve integrated transition amplitudes resulting from independent variations were calculated directly from the measured spectra and the interpolation and integration weights.

\section{Discussion and conclusion}\label{sec:discussion}
\rev{We demonstrated a target-conditioned neural compiler for the continuous family $U_2(\mathbf q)\otimes I_2\otimes I_2$ on a three-spin NMR processor. With fidelity-informed training, the nominal model attains a held-out mean unitary fidelity of approximately $0.993$ and generates controls in approximately $0.44\,\mathrm{ms}$ after training. A sensitivity diagnosis identified RF-gain error as the dominant tested uncertainty, and per-gate CVaR fine-tuning improved tolerance even outside the $\pm5\%$ training interval. Robust GRAPE remains stronger in offline uncertainty-tested fidelity; the neural advantage is amortized inference, not a demonstrated minimum equal-fidelity optimization cost.}

\rev{Hardware spectra confirm the expected systematic phase dependence of the compiled rotation family. For the Hadamard test, the measured deviation-matrix correlation drops to $0.402$ and $0.369$ at $-10\%$ and $+10\%$ RF error with nominal controls, whereas the robust controls maintain $0.774$ and $0.781$, respectively.}

\rev{The main limitation is the scaling cost of the underlying model. Dense $n$-qubit dynamics require operators of dimension $2^n$, as a result, the offline training cost would also scale exponentially. The present model is restricted to local single-qubit targets on spin 1. Extending it to parameterized entangling targets would require a wider target representation, verified controllability, and scalable propagation, for example with sparse or tensor-network methods. Hardware-calibrated uncertainty models, initial-state-aware simulations, open-system pulse propagation, and full process tomography are natural next steps.}

\appendix
\section{Backpropagation through the risk-aware neural compiler}\label{app:backprop}
This appendix writes out the gradient that is evaluated automatically during robust fine-tuning.  The derivation is included to make clear which quantities are differentiated, how the empirical CVaR term selects scenarios, and how the quantum-dynamical gradient is propagated back to the neural weights.  The implementation uses automatic differentiation, but the graph is mathematically equivalent to the chain of Jacobians below.

\subsection{Computational graph and notation}
For a target quaternion $\mathbf q_b$ and network parameters $\theta$, the robust training graph is
\begin{equation}
\mathbf q_b\xrightarrow{f_\theta}\mathbf z_b
\xrightarrow{\mathcal B}\mathbf u_b
\xrightarrow{\mathcal C_{\xi_s}}\widetilde{\mathbf u}_{b,s}
\xrightarrow{\mathcal P_{\xi_s}}U_{b,s}
\xrightarrow{F} \ell_{b,s}
\xrightarrow{\mathrm{mean/CVaR}}\mathcal L_{\rm rob},
\label{eq:Agraph}
\end{equation}
where $b=1,\ldots,B$ indexes target gates, $s=1,\ldots,S$ indexes uncertainty scenarios, $N=300$ is the number of control slices, $\mathbf z_{b,k}\in\mathbb R^2$ is the raw network output at slice $k$, $\mathcal B$ is the radial amplitude map, and $\mathcal P_{\xi_s}$ is the scenario-conditioned quantum propagation. During the reported $\pm5\%$ risk-aware training run, $B=4$ gates per microbatch and $S=10$ scenarios (eight random joint scenarios plus the two exact RF-gain endpoints). Four independently sampled microbatches are accumulated before each optimizer update.

The random target quaternions, uncertainty samples, dropout masks, and boundary scenarios are treated as fixed inputs for the backward pass in which they were drawn.  Gradients are therefore taken with respect to $\theta$ through the controls and the quantum dynamics, not with respect to the random-number generator or the physical uncertainty variables themselves.

\subsection{\texorpdfstring{\rev{Neural output map and its control Jacobian}}{Neural output map and its control Jacobian}}
\rev{The multilayer perceptron uses the architecture specified in Table~\ref{tab:nominal_hparams}. Standard network-layer Jacobians and optimizer updates are handled by automatic differentiation; the model-specific derivative needed to connect the bounded control to the physical propagation is given below.}

The bounded physical control is
\begin{equation}
\mathbf u_k
=u_{\max}\frac{\mathbf z_k}{\sqrt{1+\mathbf z_k^{\mathsf T}\mathbf z_k}},
\qquad u_{\max}=50\,\mathrm{kHz}.
\label{eq:Aradial}
\end{equation}
Writing $r_k^2=\mathbf z_k^{\mathsf T}\mathbf z_k$, the $2\times2$ Jacobian of this map is
\begin{equation}
J_{\mathcal B,k}
=u_{\max}\left[
\frac{I_2}{\sqrt{1+r_k^2}}
-\frac{\mathbf z_k\mathbf z_k^{\mathsf T}}{(1+r_k^2)^{3/2}}
\right].
\label{eq:AradialJac}
\end{equation}
Hence an upstream control gradient $\mathbf g_{u,k}=\partial\mathcal L/\partial\mathbf u_k$ becomes
\begin{equation}
\frac{\partial\mathcal L}{\partial\mathbf z_k}
=J_{\mathcal B,k}^{\mathsf T}\mathbf g_{u,k}.
\label{eq:Abackradial}
\end{equation}
The map is smooth everywhere, including at $\mathbf z_k=0$, and avoids the zero derivative introduced by hard clipping outside an admissible interval.

\subsection{Scenario-conditioned control and Hamiltonian derivative}
For scenario $s$, define
\begin{equation}
C_s=g_s
\begin{bmatrix}
\cos\phi_s&-\sin\phi_s\\
\sin\phi_s& \cos\phi_s
\end{bmatrix},
\qquad
\widetilde{\mathbf u}_{b,s,k}=C_s\mathbf u_{b,k}.
\label{eq:AscenarioMap}
\end{equation}
The drift Hamiltonian $H_{0,s}$ contains the sampled common $B_0$, spin-specific chemical-shift, and coupling perturbations, while the sampled clock error and timing jitter define $\widetilde{\Delta t}_{s,k}$.  Conditional on the sampled scenario,
\begin{equation}
H_{b,s,k}=H_{0,s}
+\pi\sum_{\beta\in\{x,y\}}
\widetilde u_{b,s,k,\beta}H_\beta.
\end{equation}
For an unperturbed commanded quadrature $u_{b,k,a}$, $a\in\{x,y\}$,
\begin{equation}
\frac{\partial H_{b,s,k}}{\partial u_{b,k,a}}
=\pi\sum_{\beta\in\{x,y\}}(C_s)_{\beta a}H_\beta.
\label{eq:AHderivative}
\end{equation}
Thus the same network control receives different gradients from different physical scenarios because the scenario map and the resulting propagator are different.

\subsection{Fr\'echet derivative of the slice propagator}
Define
\begin{equation}
A_{b,s,k}=-i\widetilde{\Delta t}_{s,k}H_{b,s,k},
\qquad
U_{b,s,k}=e^{A_{b,s,k}}.
\end{equation}
For an arbitrary perturbation $E$, the Fr\'echet derivative of the matrix exponential is
\begin{equation}
\mathcal L_{\exp}(A,E)
=\int_0^1 e^{(1-\tau)A}E e^{\tau A}\,d\tau.
\label{eq:Afrechet}
\end{equation}
Therefore
\begin{equation}
\frac{\partial U_{b,s,k}}{\partial u_{b,k,a}}
=\mathcal L_{\exp}\!\left(
A_{b,s,k},
-i\widetilde{\Delta t}_{s,k}
\frac{\partial H_{b,s,k}}{\partial u_{b,k,a}}
\right).
\label{eq:AsliceDeriv}
\end{equation}
This is the exact differential of the matrix exponential; the implementation obtains the same vector-Jacobian product through the automatic-differentiation rule.

For the ordered product
\begin{equation}
U_{b,s}=U_{b,s,N}\cdots U_{b,s,1},
\end{equation}
define the forward and backward products
\begin{equation}
P_{b,s,k}=U_{b,s,k}\cdots U_{b,s,1},\quad P_{b,s,0}=I,
\qquad
Q_{b,s,k}=U_{b,s,N}\cdots U_{b,s,k+1},\quad Q_{b,s,N}=I.
\end{equation}
The product rule gives
\begin{equation}
\frac{\partial U_{b,s}}{\partial u_{b,k,a}}
=Q_{b,s,k}
\frac{\partial U_{b,s,k}}{\partial u_{b,k,a}}
P_{b,s,k-1}.
\label{eq:AtotalDeriv}
\end{equation}
Equation~\eqref{eq:AtotalDeriv} is the time-ordered analogue of the forward/backward products that appear in GRAPE; here its control gradient is subsequently chained through a shared neural compiler rather than used to update each time-slice amplitude as an independent optimization variable.

\subsection{Trace-fidelity gradient}
For target $V_b=U_{\mathrm{tar}}(\mathbf q_b)$ and dimension $d=8$, write
\begin{equation}
c_{b,s}=\operatorname{Tr}(V_b^\dagger U_{b,s}),
\qquad
F_{b,s}=\frac{|c_{b,s}|^2}{d^2},
\qquad
\ell_{b,s}=1-F_{b,s}.
\end{equation}
For a real scalar control variable $x$,
\begin{equation}
\frac{\partial F_{b,s}}{\partial x}
=\frac{2}{d^2}\operatorname{Re}\!\left[
 c_{b,s}^{*}
 \operatorname{Tr}\!\left(V_b^\dagger
 \frac{\partial U_{b,s}}{\partial x}\right)
\right].
\label{eq:Afidgrad}
\end{equation}
Combining Eqs.~\eqref{eq:AHderivative}--\eqref{eq:Afidgrad} gives the explicit scenario-conditioned control gradient
\begin{align}
\frac{\partial \ell_{b,s}}{\partial u_{b,k,a}}
=&-\frac{2}{d^2}\operatorname{Re}\Bigg[
 c_{b,s}^{*}\operatorname{Tr}\Bigg(
 V_b^\dagger Q_{b,s,k}\nonumber\\[-1mm]
&\times\mathcal L_{\exp}\!\left(
 A_{b,s,k},
 -i\widetilde{\Delta t}_{s,k}\pi
 \sum_{\beta}(C_s)_{\beta a}H_\beta
 \right)P_{b,s,k-1}
 \Bigg)\Bigg].
\label{eq:Acontrolgrad}
\end{align}
The nominal gradient is the special case with the nominal Hamiltonian, $C_s=I_2$, and $\widetilde{\Delta t}_{s,k}=\Delta t$.

\subsection{Mean and CVaR gradient aggregation}
For one microbatch, the three fidelity terms are
\begin{align}
\mathcal L_0&=\frac1B\sum_{b=1}^{B}\ell_{b,0},\\
\mathcal L_\mu&=\frac1{BS}\sum_{b=1}^{B}\sum_{s=1}^{S}\ell_{b,s},\\
\mathcal L_{\rm tail}&=\frac1B\sum_{b=1}^{B}\frac1k
\sum_{s\in\mathcal W_k(b)}\ell_{b,s},
\label{eq:Atail}
\end{align}
where $k=\lceil\alpha S\rceil$ and $\mathcal W_k(b)$ contains the $k$ largest losses for target $b$.  In the reported risk-aware training run, $\alpha=0.20$, $S=10$, and therefore $k=2$.  Away from exact ties in the ordered losses, the empirical CVaR derivative is simply
\begin{equation}
\nabla_\theta\mathcal L_{\rm tail}
=\frac1{Bk}\sum_b\sum_{s\in\mathcal W_k(b)}
\nabla_\theta\ell_{b,s}.
\label{eq:Atailgrad}
\end{equation}
The \texttt{topk} operation is piecewise differentiable: only the selected worst scenarios receive gradient.  At an exact tie the function is nonsmooth and a valid subgradient is selected by the library's tie-breaking order.  With continuously sampled perturbations such ties have probability zero except for accidental symmetries; this does not require a differentiable relaxation of the ranking operation.

With weights $(\lambda_0,\lambda_\mu,\lambda_{\rm tail})=(1,1,1.5)$, the control-level gradient before the neural chain rule is
\begin{align}
\frac{\partial\mathcal L_{\rm rob}}{\partial\mathbf u_{b,k}}
=&\frac{\lambda_0}{B}\frac{\partial\ell_{b,0}}{\partial\mathbf u_{b,k}}
+\frac{\lambda_\mu}{BS}\sum_s
\frac{\partial\ell_{b,s}}{\partial\mathbf u_{b,k}}\nonumber\\
&+\frac{\lambda_{\rm tail}}{Bk}
\sum_{s\in\mathcal W_k(b)}
\frac{\partial\ell_{b,s}}{\partial\mathbf u_{b,k}}
+\lambda_s\frac{\partial\mathcal R_s}{\partial\mathbf u_{b,k}}
+\lambda_p\frac{\partial\mathcal R_p}{\partial\mathbf u_{b,k}}.
\label{eq:Arobcontrolgrad}
\end{align}
This equation makes explicit why tail-risk training differs from simply increasing the number of mean-robust scenarios: the worst two scenarios for each target receive an additional gradient weight on top of their contribution to the mean term.

\subsection{Regularizer gradients}
For
\begin{align}
\mathcal R_s&=\frac{1}{B(N-1)u_{\max}^2}
\sum_{b=1}^{B}\sum_{k=1}^{N-1}
\|\mathbf u_{b,k+1}-\mathbf u_{b,k}\|_2^2,\\
\mathcal R_p&=\frac{1}{BNu_{\max}^2}
\sum_{b=1}^{B}\sum_{k=1}^{N}\|\mathbf u_{b,k}\|_2^2,
\end{align}
the power gradient is
\begin{equation}
\frac{\partial\mathcal R_p}{\partial\mathbf u_{b,k}}
=\frac{2\mathbf u_{b,k}}{BNu_{\max}^2}.
\end{equation}
For an interior slice $1<k<N$,
\begin{equation}
\frac{\partial\mathcal R_s}{\partial\mathbf u_{b,k}}
=\frac{2(2\mathbf u_{b,k}-\mathbf u_{b,k-1}-\mathbf u_{b,k+1})}
{B(N-1)u_{\max}^2},
\label{eq:Asmoothgrad}
\end{equation}
with the obvious one-sided endpoint forms.  The smoothness term is therefore a discrete second-difference penalty in the backward pass.

\subsection{\texorpdfstring{\rev{Network gradient accumulation and update}}{Network gradient accumulation and update}}
\rev{The control gradient in Eq.~\eqref{eq:Arobcontrolgrad} is pulled back through the radial map by Eq.~\eqref{eq:Abackradial} and then through the shared network by automatic differentiation. Four independently sampled microbatch gradients are averaged before each parameter update; the global gradient norm is clipped to $1$ and the network parameters are updated with AdamW (default moment coefficients $\beta_1=0.9$, $\beta_2=0.999$, numerical $\epsilon=10^{-8}$, decoupled weight decay $10^{-3}$). The robust fine-tuning learning-rate schedule starts at $1.5\times10^{-4}$. These operations use standard neural-network optimization rules; the scenario-dependent quantum-dynamical gradients and per-gate CVaR selection are the method-specific elements derived above.}

\subsection{Interpretation as amortized robust optimal control}
The backpropagation can be read in two stages.  Equations~\eqref{eq:AsliceDeriv}--\eqref{eq:Acontrolgrad} compute the same type of time-ordered physical sensitivity used by gradient-based quantum optimal control.  \rev{Equation~\eqref{eq:Arobcontrolgrad} aggregates those sensitivities across targets and uncertainty scenarios, and Eq.~\eqref{eq:Abackradial} carries them into the shared network $f_\theta$.}  GRAPE updates $600$ control coordinates independently for one target; the compiler updates shared parameters so that one forward pass produces all $600$ coordinates for any target quaternion.  Robust CVaR training adds another coupling: for each target, scenarios in the current adverse tail receive extra gradient weight.  This is the precise mathematical sense in which the reported model performs amortized, tail-risk-aware quantum optimal control.

\section*{Acknowledgments}

During manuscript preparation, the authors used OpenAI ChatGPT to assist with language editing, clarity, and presentation. The authors reviewed and revised all AI-assisted output and take full responsibility for the scientific content, analysis, interpretation, and conclusions.

\section*{Statements and Declarations}

\subsection*{Funding}

This work is supported by Koç University.

\subsection*{Competing interests}

Zidong Lin is affiliated with Shenzhen SpinQ Technology Co., Ltd. The experimental methodology and conclusions of this work do not rely on proprietary features specific to SpinQ hardware; the reported control and tomography procedures can, in principle, be implemented on comparable benchtop NMR spectrometers. The remaining authors declare no relevant financial or non-financial interests related to the content of this work.

\subsection*{Author contributions}

Arash Fath Lipaei conceived the idea of robustness and co-conceived the idea of neural-network-based pulse generation, developed and implemented the numerical framework for pulse generation, training, and testing, wrote the mathematical formulation and training-methodology sections and reviewed the manuscript. Ebrahim Khaleghian co-conceived the idea of neural-network-based pulse generation, reviewed and validated the numerical work, examined and verified the physical modeling of the device, wrote the physics-related sections of the manuscript, implemented the experimental component and tomography-reconstruction code, verified the experimental results, and contributed to writing and reviewing the manuscript. Gani Göral facilitated access to the experimental setup and reviewed the manuscript. Zidong Lin verified the physical scaling and parameter conversion used for the experimental device. Selin Aslan supervised the research, contributed to the reviewing of the manuscript and reviewed and verified the methodology, results, and interpretation. Özgür E. Müstecaplıoğlu supervised the research, contributed to the writing and the reviewing of the manuscript and reviewed and verified the methodology, results, and interpretation. All authors reviewed and approved the final manuscript. Arash Fath Lipaei and Ebrahim Khaleghian contributed equally to this work and share first authorship.

\section*{Data and code availability}
The processed data, numerical outputs, digitized tomography spectra, reconstruction code, and analysis scripts supporting the findings of this study are publicly available in the accompanying reproducibility repository, \texttt{Robust-NeuralQControl}, at \href{https://github.com/Ar2000Ash/Robust-NeuralQControl}{https://github.com/Ar2000Ash/Robust-NeuralQControl}.

\begingroup
\setlength{\bibsep}{0pt}

\bibliographystyle{plainnat}
\bibliography{abcd}

\endgroup

\end{document}